# Direct Visualization of Native Defects in Graphite and Their Effect on the Electronic Properties of Bernal-Stacked Bilayer Graphene

Frédéric Joucken,[1,2,†] Cristina Bena,[3,†] Zhehao Ge,[1] Eberth Quezada-Lopez,[1] Sarah Pinon,[3] Vardan Kaladzhyan,[4] Takashi Taniguchi,[5] Kenji Watanabe,[6] Aires Ferreira,[7] Jairo Velasco Jr.[1,†]

[1]*Department of Physics, University of California, Santa Cruz, California, USA*

[2]*Department of Physics, Arizona State University, Tempe, USA*

[3]*Institut de Physique Théorique, Université Paris Saclay, CEA CNRS, Orme des Merisiers, 91190 Gif-sur-Yvette Cedex, France*

[4]*Department of Physics, University of Basel, Klingelbergstrasse 82, CH-4056 Basel, Switzerland*

[5]*International Center for Materials Nanoarchitectonics, National Institute for Materials Science, 1-1 Namiki, Tsukuba 305-0044, Japan*

[6]*Research Center for Functional Materials, National Institute for Materials Science, 1-1 Namiki, Tsukuba 305-0044, Japan*

[7]*Department of Physics and York Centre for Quantum Technologies, University of York, York YO10 5DD, United Kingdom*

[†]Email: frederic.joucken@gmail.com, cristina.bena@ipht.fr, jvelasc5@ucsc.edu

**Abstract**

Graphite crystals used to prepare graphene-based heterostructures are generally assumed to be defect free. We report here scanning tunneling microscopy results that show graphite commonly used to prepare graphene devices can contain a significant amount of native defects. Extensive scanning of the surface allows us to determine the concentration of native defects to be $6.6 \times 10^8$ cm$^{-2}$. We further study the effects of these native defects on the electronic properties of Bernal-stacked bilayer graphene. We observe gate-dependent intravalley scattering and successfully compare our experimental results to T-matrix-based calculations, revealing a clear carrier density dependence in the distribution of the scattering vectors. We also present a technique for evaluating the spatial distribution of short-scale scattering. A theoretical analysis based on the Boltzmann transport equation predicts that the dilute native defects identified here are an important extrinsic source of scattering, ultimately setting the mobility at low temperatures.

**Introduction**

In the early days of graphene research, when transport experiments were carried out on graphene supported by $SiO_2$, significant attention was devoted to understanding the effects of charge impurities in graphene samples. Notably, the minimum conductivity was measured to be close to twice the quantum of conductance ($4e^2/h$) in zero magnetic field.[1–3] This was shown to be an extrinsic property of graphene resulting from the inhomogeneous potential landscape created by charged impurities in the supporting $SiO_2$.[4,5] Later, demonstration of ballistic transport in suspended graphene samples[6,7] and in graphene encapsulated in hexagonal boron nitride (hBN),[8] shifted the attention away from impurity-induced scattering in graphene samples. For suspended samples, a consensus on the primary role of electron-phonon interaction as limiting the mobility seems to have emerged.[9,10] For hBN supported samples, magnetotransport experiments on mono-[11] and Bernal-stacked bilayer graphene (BLG)[12] have pointed towards intravalley scattering as the dominant factor limiting the electronic mobility on this substrate. The intravalley scattering was attributed to strain-induced pseudomagnetic fields.[11,12] More recently however, the role of localized impurities as scatterers has reemerged in the context of twisted graphene samples, with indications that they might play a key role in these systems.[13–15]

Interestingly, to the best of our knowledge, there has been so far no report of scanning tunneling microscopy (STM) study of *native* point-like defects in graphite used for making graphene devices. Trenches and steps[16–18] as well as grain boundaries[18–20] have been investigated by STM but these types of defects are easily avoidable when making devices from exfoliated graphene flakes as they are visible under an optical or an atomic force microscope. Several STM studies have focused on point-like defects in graphite but all of these were obtained on samples with intentionally induced defects (mostly by ion irradiation).[21,22] As for graphene, several investigations of native point-like defects have been reported[23–26] but they were not performed on exfoliated graphene and the influence of these defects on the transport properties was not addressed. The only atomic scale study carried out on exfoliated graphene that focused on the role of localized charged impurities was by Zhang et al.[27] It revealed that the impurities were not defects in the graphene lattice but likely molecules trapped between graphene and the supporting $SiO_2$, a substrate that has become nearly obsolete for electronic devices since the advent of hBN for this role. The absence of direct experimental evidence for the presence of native defects in graphene devices and in graphite parent crystals likely explains why the role of defects in the transport properties of graphene devices has been thus far mostly overlooked.

Here we report scanning tunneling microscopy (STM) results that establish the presence of atomic-scale defects in a type of graphite commonly used to make graphene devices with the standard exfoliation technique. We further characterize the effects of these defects on the electronic properties of BLG. By mapping quasiparticle scattering interference (QPI) on mesoscopic areas (typically $> 300 \times 300$ nm²) we visualize the intravalley scattering patterns induced by quasiparticles scattered off localized defects. We compare our experimental QPI results to T-matrix-based calculations and reveal the dramatic influence of the charge carrier density and the perpendicular electric field (tuned by the gate voltage) on the distribution of the scattering vectors. We also study the spatial extension of intervalley scattering induced by the same localized defects and show that it extends significantly ($> 10$ nm) away from them, following patterns that resemble the intravalley scattering patterns produced around the same defects. Subtle differences between the inter- and the intra-valley patterns are however observed and explained. Finally, we present a theoretical analysis based on the Boltzmann transport equation that strongly indicates that, despite their low concentration, the native defects we have observed become the dominant source of scattering at low temperature.

Several factors enabled our discovery of native defects in graphite crystals that are commonly used for mechanical exfoliation but were so far unreported. It was important to image numerous large areas (typically $> 200 \times 200$ nm²) at low tip-sample bias (few mVs) and at relatively high current (typically 1-2 nA; sometimes greater, depending on the tip state). For the experiments performed on devices, using a BLG sample enhanced the visibility of the QPI patterns compared to the monolayer case. We attribute this to the scattering selection rules being less favorable to producing visible patterns for monolayer graphene.[28,29]

**Native defects in graphite**

Figure 1a shows a typical large-scale ($500 \times 500$ nm²) low-bias (4 mV) STM topographic image of a freshly exfoliated graphite crystal ("Flaggy Flakes" from NGS Naturgraphit; we obtained similar results on "Graphenium Flakes" from the same company). On this image, several localized triangular patterns can be distinguished. By zooming in on each of these patterns and acquiring atomically resolved images with our STM, we found that 20-30% of the triangular patterns observed on large scale images had a localized atomic-scale defect located at their center, in the topmost layer. Such an atomic scale defect is shown in the inset of Fig. 1 (other representative defects are shown in the SM, Section 1). The nature of most of these defects remains unknown, but we tentatively identify a few of them as nitrogen dopants (see SM Section 1). The 20-30% of triangular patterns that were found to have an atomic scale defect at their center (in the topmost layer) appeared the brightest on the large-scale images. From this, we conclude that 70-80% of the patterns that appear on large scale images are produced by atomic scale defects located in buried layers. By counting the total number of defects found on the topmost layer of freshly exfoliated graphite, we determine the defect concentration to be $(6.6 \pm 2) \times 10^8$ cm$^{-2}$ *per layer* (see SM Section 2 for details on the counting and uncertainty). Such a concentration of native defects, albeit seemingly low, makes it challenging to fabricate devices which are defect free. We plot in Fig. 1b the probability of having no defect in a monolayer as a function of the area, assuming a uniform and random distribution of defects and considering the concentration we have measured (details in SM, Section 2). One can see that to have a reasonable chance of attaining a defect-free monolayer sheet, areas smaller than ~ 0.2 µm² should be considered. This becomes even more stringent when considering multilayer samples.

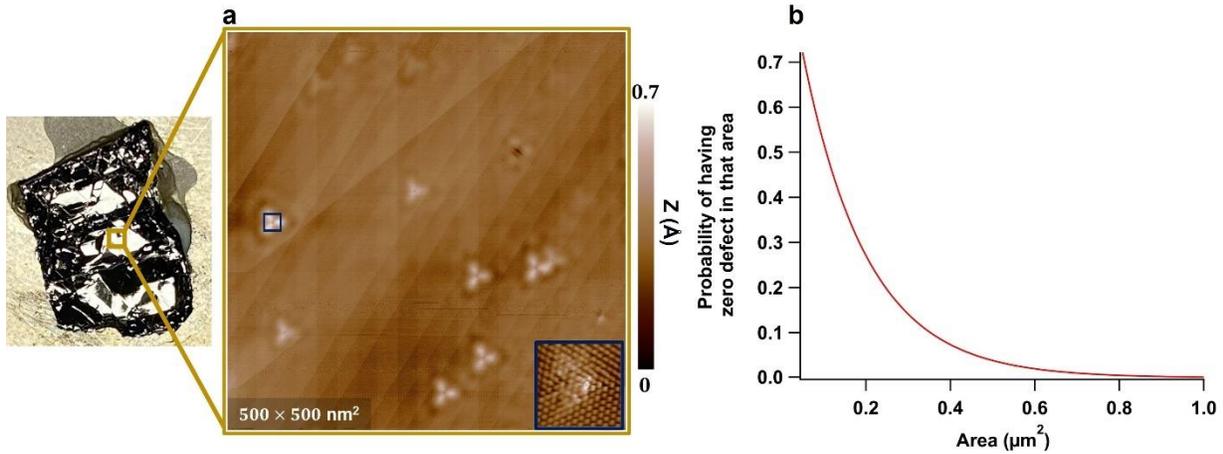

*Fig. 1: **Native defects in graphite.** (a) Large-scale ($500 \times 500$ nm²) low-bias (4 mV) STM topographic image of graphite ("Flaggy Flakes" from NGS Naturgraphit), exfoliated in UHV moments prior to the STM experiments. Quasiparticle interference (QPI) patterns with triangular symmetry are visible. As discussed in the text, these QPI patterns are attributed to scattering off atomic scale defects. A zoom in around one of these defects (blue square) located in the topmost layer is shown in the inset. (b) Probability of having no defect in a monolayer, as a function of the area, assuming a defect concentration of $(6.6 \pm 2) \times 10^8$ cm$^{-2}$.*

Having established the presence of native defects in the graphite parent crystal, we now move on to study the effects these defects impart on the electronic properties of a back-gated BLG device made from the same parent crystal shown in Fig. 1a.

**Native defects in BLG**

We assembled BLG/hBN heterostructures that were deposited on SiO$_2$/Si (see methods), with the doped silicon substrate serving as an electrostatic gate; a schematic of the device is shown in Fig. 2a. Before analyzing the scattering induced by the dopants in our BLG/hBN heterostructure, we point out that we measured a defect concentration in graphene/hBN devices of $(2.2 \pm 0.7) \times 10^9$ cm$^{-2}$ per layer (see SM Section 2 for details on the counting and uncertainty; graphene devices include bilayer and multilayer graphene), significantly greater than the defect concentration measured on the freshly exfoliated graphite parent crystal. Thus, it appears that our sample making procedure (see methods), which involves heating the graphene on hBN stacks in forming gas (Ar/H$_2$) as well as prolonged annealing of the devices in ultra-high vacuum at 400 °C results in the creation of additional defects. These two annealing steps are commonly used by researchers studying graphene samples in UHV. We also note that for the BLG/hBN sample, we have counted a number of defects on the top layer which was somewhat smaller than half of the total number of defects (determined by counting the large-scale intravalley scattering patterns). This suggests that defects in the hBN substrate[30] or impurities trapped between the hBN and the BLG might be responsible for some of the intravalley scattering patterns that we have observed.

To our knowledge, the only other report of imaging native defects in graphene devices was by Halbertal et al.,[31] who have reported much smaller native defects concentration than what we report here (they saw 3 defects in a $4 \times 4$ μm² area, corresponding to $1.9 \times 10^7$ cm$^{-2}$). We note however that the technique they used (scanning nanothermometry on encapsulated graphene) is less direct than STM and the effect they attributed to atomic scale defects could have been produced by defects complexes or trapped impurities. We also note that the graphite parent crystal used was not indicated in their study.

**Intravalley scattering**

Figure 2b shows a topographic STM image of a large (515 × 290 nm²) area. Figures 2c-i show dI/dV$_S$ maps acquired in the same region as shown in Fig. 2b, at various gate voltages (indicated), together with their corresponding fast Fourier transform (FFT). Because the tip-sample bias was low (+5 mV), these maps essentially reveal the local density of states at the Fermi level (LDOS($E_F$)). The difference between $E_F$ and the charge neutrality point ($E_{CNP}$) is indicated (see SM Section 3 for details). Clear QPI patterns (also known as Friedel oscillations) are seen on each of these dI/dV$_S$ maps. Importantly, these QPI patterns visibly originate from localized scattering centers, as is particularly evident at low gate voltage (panels c and f of Fig. 2). The QPI patterns reported in Fig. 2 correspond to intravalley scattering (as schematized in the inset of Fig. 2c). This appears unambiguously in the FFTs, where the size of the QPI patterns (centered around the origin) have dimensions that correspond to the size of the Fermi surface at the corresponding gate.[32] The hexagonal shape of the patterns seen in the FFT is due to the strong trigonal warping of the low energy bands in bilayer graphene.[32] We note that the FFT amplitude is for most cases close to zero within the hexagonal boundary of the scattering pattern and maximum at the boundary, except for low gate voltages ($V_G = +20$ V and $V_G = -10$ V) where significant intensity is observed for short scattering vectors (small momentum transfer). We also clearly see that the overall amplitude of the FFTs is greater for positive gate voltages than for negative gate voltages (all the FFTs have the same z-scales).

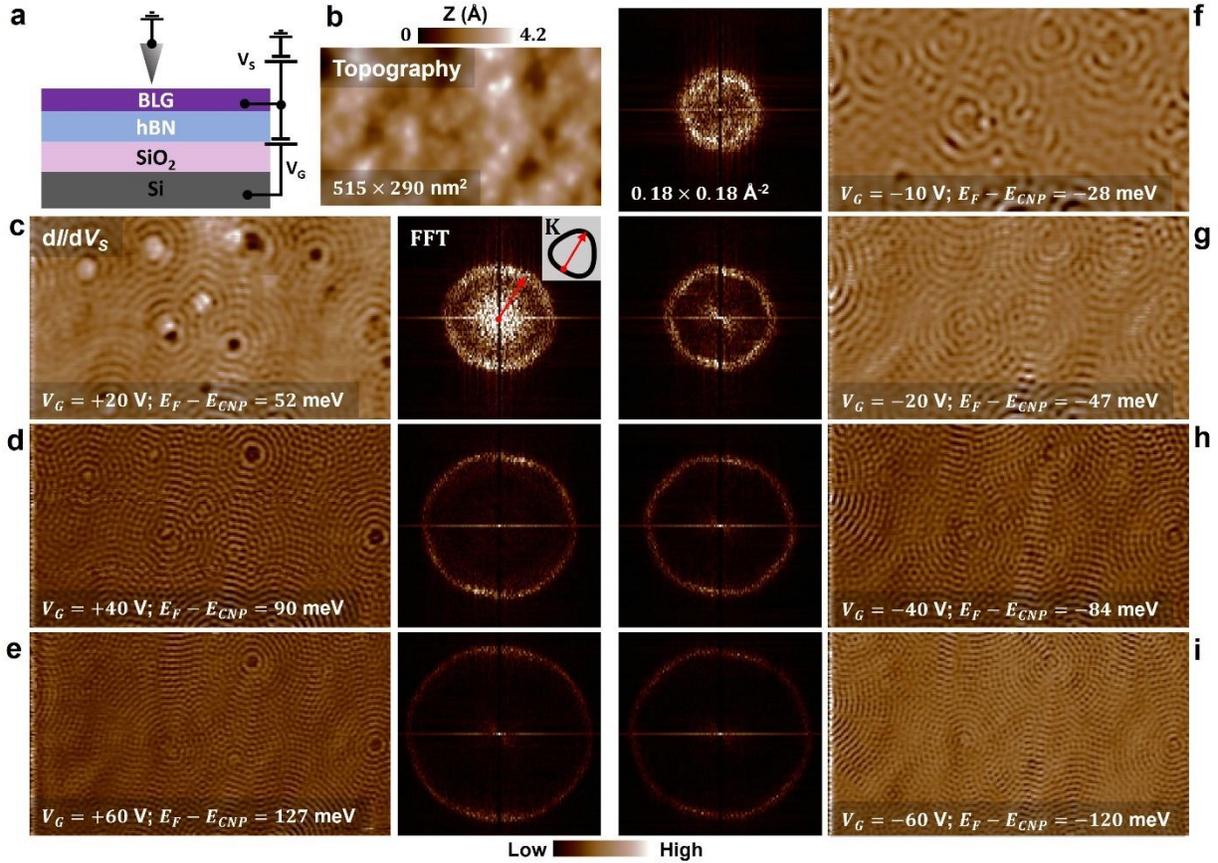

*Fig. 2:* **Intravalley scattering in gate tunable Bernal-stacked bilayer graphene.** *(a) Schematics of the experimental setup. A back-gated bilayer graphene flake atop a ~20nm-thick hBN flake is scanned with the STM. (b) Large-scale (515 × 290 nm²) topographic STM image of a clean bilayer grahene area. (c)-(i) dI/dV$_S$ maps (3 mV excitation) acquired over the same area as in (b), at various gate voltages (V$_G$; indicated), at low sample bias (+5 mV; I=0.5 nA), essentially mapping the local density of states at the Fermi level. The difference between the Fermi level (E$_F$) and the charge neutrality point (E$_{CNP}$) is indicated for each gate voltage. The fast Fourier transform (FFT) of each map is also shown (all the FFTs are plotted on the same z-scale). Clear intravalley scattering (schematized in the inset of (c)) pattern is visible both in real space and in the FFT. Whereas all FFT patterns show stronger intensity at the edge of the pattern (empty circles), the FFT pattern at V$_G$ = +20 V displays strong intensities for small momentum transfer. This is discussed further in the text, in connection with Fig. 3.*

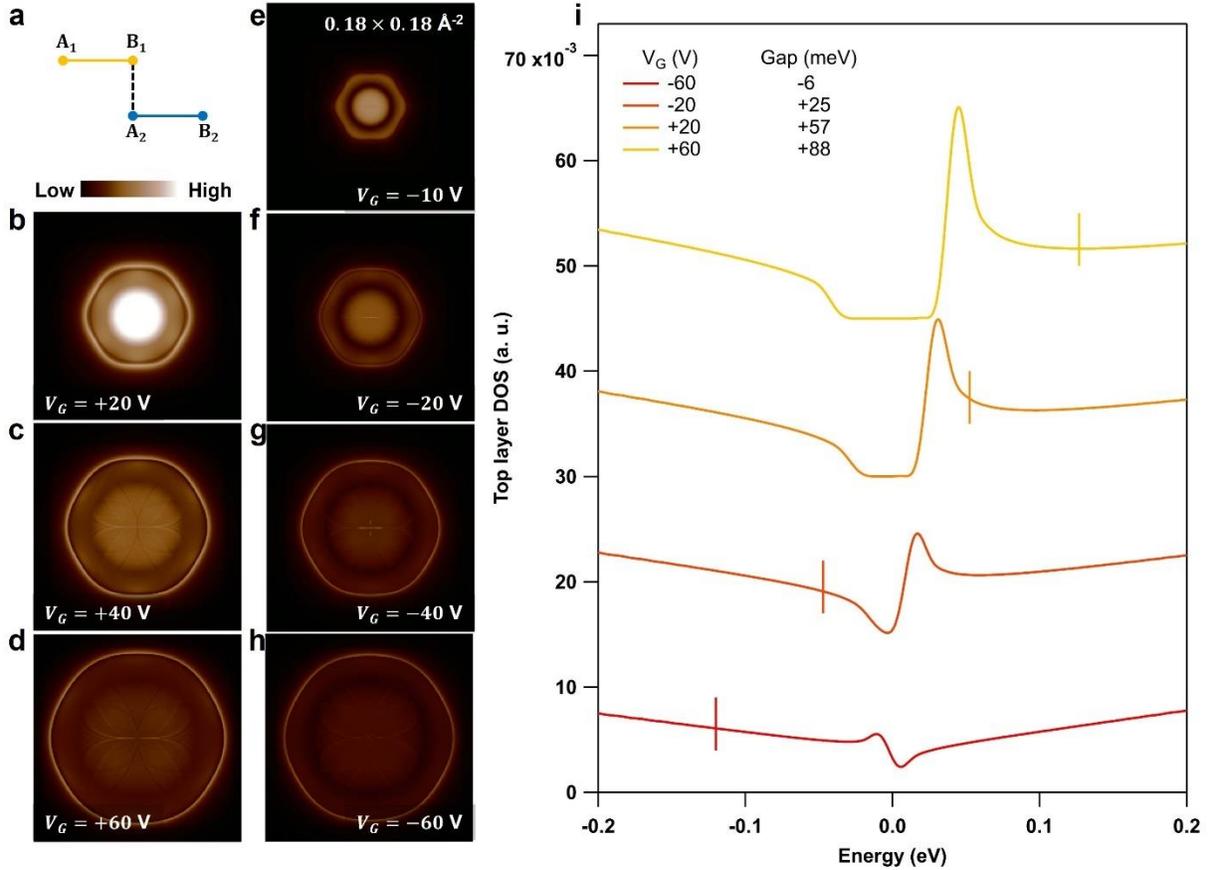

*Fig. 3: **Simulated reciprocal-space signature of defect-induced QPI in gate-tunable Bernal-stacked bilayer graphene.** (a) Sublattices in Bernal-stacked bilayer graphene unit cell. (b-h) T-matrix simulations of the QPI in momentum space, for the indicated gate voltages. (i) Top-layer DOS for the gate voltages indicated. The vertical bar indicates the energy at which the corresponding QPI was acquired (and simulated).*

Next, we examine the reciprocal-space signature of the QPI that we obtain from T-matrix simulations[28,33] and compare it to our experimental results (FFTs in Fig. 2). Figures 3b-3h show the simulated FFT for the same doping as in the experiments ($E_F - E_{CNP}$ indicated in Fig. 2), all on the same z-scale. The band gap induced by the perpendicular electric field produced by the backgate and the STM tip is considered in the simulation.[34–36] The method for determining the values used in the simulation for the band gap and for the electronic doping for each gate voltage is presented in section 3 of the SM. The T-matrix results presented in Fig. 3b-3h are obtained by averaging the signatures of dopants located on the four atomic sites of the BLG unit cell ($A_1$, $B_1$, $A_2$, and $B_2$, see inset of Fig. 3a), and only considering the DOS in the top layer (similar to the situation for the STM experiment). The dopant was modelled by an onsite potential of $-10$ eV (we discuss in the section 6 of the SM the influence of the potential on the QPI signature in real and reciprocal space). One can see that the main features of the experimental FFTs (Fig. 2 c-i) are captured by the simulations: (i) the overall amplitude of the FFTs is greater for positive gates than for negative ones; and (ii), the spectral weight within the hexagonal pattern, and particularly in the vicinity of the origin (enhancement of small momentum transfer), is greater for smaller gate voltages.

Deeper insight into these two features can be gained by looking at the calculated gate-dependent density of states (DOS) of the top layer (which is the layer probed by STM). Figure 3i displays the top-layer DOS for pristine BLG computed for four gate voltages of panels (b-h). The vertical lines indicate the corresponding position of the Fermi level at these gate voltages (and thus the energy that was probed in the experiments). The values of the gap used to compute these curves are indicated (see SM section 3 for the determination of these values). As is observed for the experimental and simulated QPI patterns, an asymmetry between positive and negative gate voltages (feature (i) of the FFTs) is clearly seen for the top-layer DOS. This asymmetry is due to the polarization of the BLG sheet in the z direction upon application of the electric field induced by the gate and the STM tip. Further asymmetry is produced by the gap closing at $V_G \approx -52$ V rather than $V_G = 0$ V. This offset is attributed to the work function mismatch between the tip and the sample[35,36].

As a tentative explanation of feature (ii) (enhancement of small momentum transfer for low gate voltages), we first note the proximity of the gap edge in these cases ($V_G = +20$ V, $V_G = -10$ V, and $V_G = -20$ V). However, the precise mechanisms behind this enhancement cannot be explained by pure band structure arguments, as we demonstrate in the SM (Section 7) by showing computed joint density of states at various gate voltages. The explanation for the enhancement of small momentum transfer at low gate voltage is thus related to the details of the scattering mechanism, such as the overlapping matrix elements between incoming and outgoing states or local modifications of the density of states. These are naturally accounted for in the T-matrix formalism.[37–40] A close examination of the dI/dV$_S$ map at $V_G = +20$ V (Fig. 2c) reveals that the scattering patterns change significantly compared to the other dI/dV$_S$ maps. The wavelike patterns emanating from the point-like scattering centers decay noticeably faster than in the maps taken at higher gate voltages. Also, a high-intensity dI/dV$_S$ signal is observed at the location of some scattering centers. Given the basic properties of Fourier transformation (recall that the Fourier transform of the sinc function is the rectangular function), these features can explain the higher intensity close to the origin in the FFT.

**Intervalley scattering**

Having thoroughly discussed the intravalley scattering caused by native defects, we now investigate the intervalley scattering induced by the same defects in our sample. The intervalley scattering induced by localized defects has already been reported in STM studies on mono- or bi-layer graphene.[23,41,42] However, the spatial extension of the intervalley scattering away from localized defects has so far lacked attention. In the case of nitrogen dopants for example, the $\sqrt{3} \times \sqrt{3}R30$ pattern associated to intervalley scattering is clearly visible in topographic images near the dopants but seems to quickly (~1 − 2 nm) decay away from the defect.[41] No technique for visualizing the spatial extent of the intervalley scattering induced by point-like defects in graphitic systems has been reported so far. We present below data and an analysis method that allow us to infer the spatial distribution of the intervalley scattering.

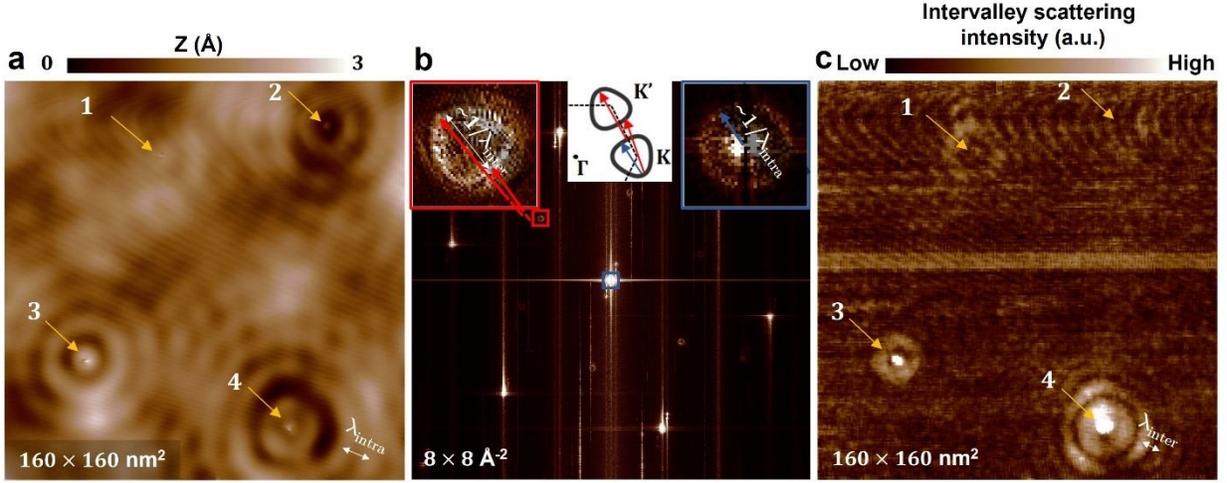

*Fig. 4: **Spatial mapping of intervalley scattering by native defects in Bernal-stacked bilayer graphene.** (a) Low-bias (5 mV) high-resolution (2048² pixels) STM topographic map of BLG. Four defects are visible (labelled 1-4). Defects 3 and 4 are in the top layer. (b) FFT of (a) where the intervalley scattering (schematized in the inset). Red-boxed inset is a zoom-in around one intervalley scattering feature; the two red arrows indicate extreme intervalley scattering vectors for this pattern. Blue-boxed inset is a zoom-in around the intravalley scattering feature; the blue arrow indicates an extreme intravalley scattering vector. The middle inset illustrates the inter- and intravalley scattering vectors. (c) Intervalley scattering map on the same area as (a) obtained by taking the ratio between the intravalley feature amplitude and the lattice feature amplitude in 2.5 × 2.5 nm² windows in image (a); see main text and SM for details on this computation.*

Fig. 4a shows a high-resolution (2048 × 2048 pixels) low-bias (+5 mV) and large-scale (160 × 160 nm²) STM topographic image. Such a map is ideal to study both inter- and intra-valley scattering because it captures both the short and long wavelengths associated with the two types of scattering. Several defects are indicated by yellow arrows in Fig. 4a. Defects 1 & 2 are buried while he defects 3 & 4 are on the top layer (see SM Section 8 for zoom-ins of STM images). The intravalley scattering induced by three of the defects (1, 3, and 4) is evident from the long wavelength patterns surrounding these defects. The intervalley scattering appears in the FFT of Fig. 4a, shown in Fig. 4b as small hollow pockets[32] (one of them is boxed in red and shown in greater detail in the inset) while the brightest spots correspond to the atomic lattice (Bragg peaks). The origin of the triangular shape of these hollow pockets is discussed in detail in ref.[32] Briefly, it is determined by the joint density of states between adjacent valley, as schematized in the middle inset of Fig. 4b, and its triangular shape is dictated by the strong trigonal warping in BLG. We also highlight in the blue boxed inset of Fig. 4b the intravalley scattering pattern. To study the spatial distribution of the intervalley scattering, we superpose a 256 × 256 grid on the image in Fig. 4a and for each point of that grid, we consider the associated 2.5 × 2.5 nm² window (corresponding to 32 × 32 pixels). For each of these windows, we then compute the FFT and evaluate the intensity ratio between the intervalley scattering features and the lattice features (further detail on this method is given in the SM, section 9).

The result of this analysis is shown in Fig. 4c. Several observations can be made. The intervalley scattering at the location of the defect is strong for defects located on the top layer (defects 3 & 4). Although both lie in the top layer, the intervalley scattering induced by defect 4 is significantly more intense than that by defect 3. In addition, the intervalley scattering at the defect location is more intense for defect 4 and it extends significantly further away from the defect than for defect 3. For defect 4, the intervalley scattering extends about 15 nm away from the defect.

Interestingly, the intensity of the intervalley scattering both for defects 3 & 4 (Fig. 4c) seems to mimic the intravalley wavelike pattern visible in Fig. 4a. This is especially visible around defect 4 in Fig. 4c, where a clear wavelike pattern with long wavelength surrounds the defect. A closer examination reveals that the wavelength of the pattern observed in the intervalley ($\lambda_{inter}$) scattering map (Fig. 4c) is two times shorter than the wavelength of the intravalley pattern ($\lambda_{intra}$) observed in Fig. 4a (~6.5 nm *vs.* ~13 nm, respectively).

The origin of the different wavelengths related to the inter- and intra-valley scattering patterns are shown in the insets of Fig. 4b. The width of both patterns are roughly equal (and equal to $4q_F$, with $q_F \approx 0.025$ Å$^{-1}$ in this case) and is determined by the corresponding extreme scattering vectors, as schematized in the middle inset of Fig. 4b.[23,29] However, in the case of intervalley scattering, the wavelength observed is λ$_{inter}$ ∼2π/4q$_F$, corresponding to the difference between the two extreme scattering vectors (see the two red vectors in the red-boxed inset in Fig. 4b). In the case of intravalley scattering, the wavelength observed is λ$_{intra}$ ∼2π/2q$_F$, corresponding to a wavevector whose length is half the intravalley pattern diameter (see the blue arrow in the blue-framed inset in Fig. 4b). This difference is due to the fact that the observed intervalley scattering is caused by beating between the two red vectors (see red-boxed inset in Fig. 4b), whereas the observed intravalley scattering is caused by the scattering vectors themselves (such as the blue vector in Fig. 4b). Further details on this are provided in the SM, section 10.

**Influence of the native dopants on the electronic transport properties of BLG**

Finally, we evaluate the implications of our discovery of native dopants in graphite for the electronic transport properties of BLG, using Boltzmann transport theory. The dopants are modeled as short-range scatterers with a typical potential strength $u = -10$ eV and areal density $n_{\text{dop}} = 10^9$ cm$^{-2}$ (per layer) equally distributed over the A-B sublattices (Fig. 3a), as per the QPI simulations above. The longitudinal dc conductivity ($\sigma_{xx}$) is obtained via the exact solution of the linearized Boltzmann transport equation within a continuum 2-band model of Bernal-stacked BLG (assuming no bulk band gap; this is reasonable since the model is valid only far from the CNP), which captures the essential aspects of electronic transport at low energies—see SM Section 11. In the zero-temperature limit, we obtain $\sigma_{xx} = \frac{2e^2}{h} v_F k_F \tau_\parallel$, with $k_F$ the Fermi wavevector, $v_F = 2k_F \beta/\hbar$ the Fermi velocity and $\tau_\parallel = \hbar \left[1 + \left(\frac{8\beta}{A_{\text{u.c.}} u}\right)^2\right]/(4n_{\text{dop}}\beta)$ the transport time. Here, $\beta = v^2 \hbar^2/t_\perp$ (with $t_\perp$ the interlayer hopping integral and $v \approx 10^6$ m/s the bare quasiparticle velocity) and $A_{\text{u.c.}}$ denotes the unit cell area. We note that $\sigma_{xx}$ has little sensitivity to thermal fluctuations provided the Fermi energy is much greater than $k_B T$. The electron mobility is given by $\mu_e = \sigma_{xx}/(en_e)$, with $n_e = \pi k_F^2$ the charge carrier density. For typical material parameters[32,43] (i.e., $\gamma_1 \approx 0.42$ eV and $A_{\text{u.c.}} \approx 1.0$ nm$^2$) we find $\mu_e \approx 3.1 \times 10^6$ cm$^2$/V/s, which is consistent with the highest reported values for hBN-encapsulated[44,45] or suspended[46–48] devices. Because the estimated mobility is one order of magnitude higher than the theoretical upper limit set by acoustic phonon scattering,[49,50] it is important to assess the crossover between impurity-scattering and intrinsic phonon-limited mobility. The mobility as a function of temperature for moderate carrier density is shown in Fig. 5, where the extrinsic and intrinsic (acoustic phonon) mobilities are compared. Our calculations show that for temperatures below 1 K, the carrier mobility is essentially limited by extrinsic scattering (from native dopants). This holds for a large window of charge carrier densities $n_e \in [10^{11}, 10^{13}]$ (cm$^{-2}$), with the crossover moving towards slightly higher temperatures in more defective samples (e.g., for $n_{dop} = 10^{11}$ cm$^{-2}$, the extrinsic mechanism is found to dominate the electronic transport for $T \lesssim 2$ K). The calculation of the phonon-limited mobility follows Ref. [49] and for completeness is outlined in SM, section 11.

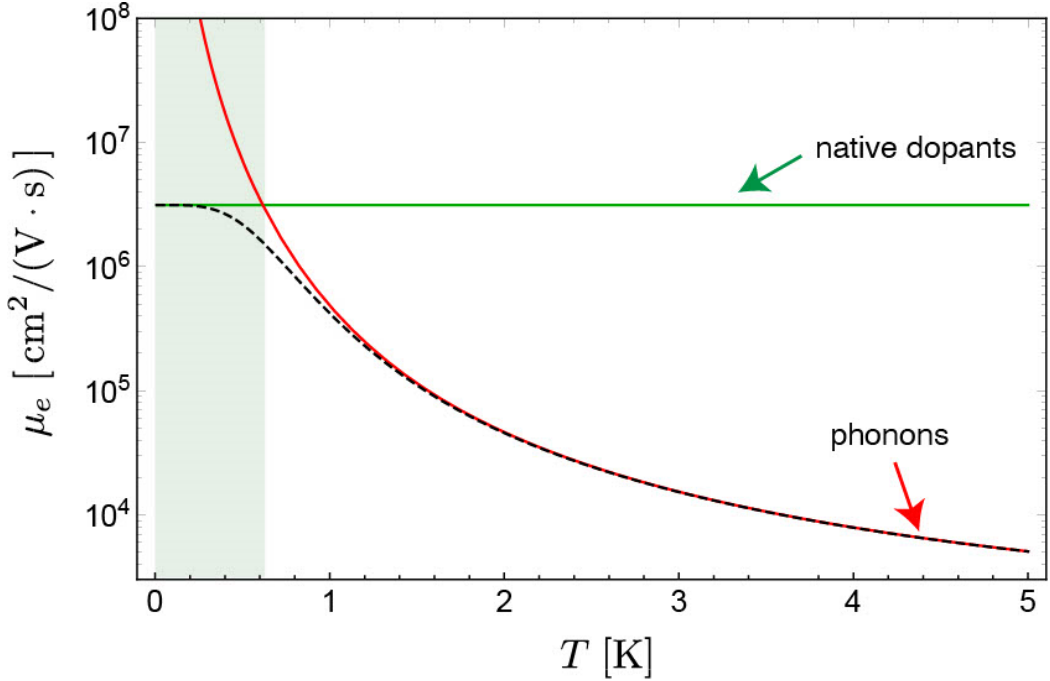

*Fig. 5: **Calculated temperature dependence of carrier mobility in Bernal-stacked bilayer graphene for a dopant concentration of $1 \times 10^9$ cm$^{-2}$**. Solid lines show the native-dopant-scattering (green) and acoustic-phonon-scattering limited (red) mobility calculated within the framework of the Boltzmann transport theory. The total charge carrier mobility estimated using Matthiessen's rule is also shown (dashed line). The charge carrier density is fixed at $n_e = 5 \times 10^{12}$ cm$^{-2}$ (for other parameters see the main text). Details on the intrinsic phonon-limited mobility calculation are given in SM, Section 11. Shaded green area indicates temperature region where scattering from native dopants is dominant.*

**Conclusion**

We have presented STM data that unambiguously demonstrates the presence of native defects in graphite commonly used to make graphene devices. We have further characterized thoroughly the effects of these defects on the electronic properties of BLG. T-matrix calculations reproduced well the gate-dependent reciprocal-space QPI signatures observed experimentally, which are not captured by simple band structure arguments. We have also presented an original method for studying the spatial dependence of intervalley scattering. We have further presented Boltzmann transport calculations predicting that the conductivity of BLG at low temperature (<1K) and moderate to high carrier density is limited by extrinsic scattering induced by native dopants. We expect that the effects discussed in this work will play an important role in other graphene systems, including flatbands systems (bilayer, trilayer, double bilayer, etc.).[51–58] Extending the methods presented here to these systems should readily reveal the effects of native defects on the exotic correlated states in the magic-angle systems.

**Methods**

Sample Fabrication

The graphite ("Flaggy Flakes" and "Graphenium Flakes" from NGS Naturgraphit) sample was exfoliated *in situ* and introduced in the STM head within seconds after the exfoliation. The graphene (bilayer and multilayers) heterostructures were stacked on hBN using a standard polymer-based transfer method.[59] A graphene flake exfoliated on a methyl methacrylate (MMA) substrate was mechanically placed on top of a 20 − 50 nm thick hBN flake that rests on a SiO2/Si++ substrate where the oxide is 285 nm thick. Subsequent solvent baths dissolve the MMA scaffold. After the Graphene/hBN heterostructure is assembled, an electrical contact to graphene is made by thermally evaporating 7 nm of Cr and 200 nm of Au using a metallic stencil mask. The single-terminal device is then annealed in forming gas (Ar/$H_2$) for six hours at 400 °C to reduce the amount of residual polymer left after the graphene transfer. To further clean the surface of the sample, the heterostructure is mechanically cleaned using an AFM.[60,61] Finally, the heterostructure is annealed under UHV at 400 °C for seven hours before being introduced into the STM chamber.

STM Measurements

The STM measurements were conducted in ultra-high vacuum with pressures better than $1 \times 10^{-10}$ mbar at 4.8 K in a Createc LT-STM. The bias is applied to the sample with respect to the tip. The tips were electrochemically-etched tungsten tips, which were calibrated against the Shockley surface state of Au(111) prior to measurements. The STM images presented in the main text and their FFTs were treated with Igor Pro. WSxM[62] was also used for data presented in the SM.

**Associated Content**

This material is available free of charge via the internet at http://pubs.acs.org.

I. Atomic-scale images of defects in exfoliated graphite and in graphene devices; II. Concentration of defects; III. Extraction of the gap and the electronic doping for the BLG device for each gate voltage; IV. Real space imaging of native defects in BLG/hBN devices; V. STM topographic map and dI=dVS maps at lower current; VI. Dependence of the real-space and reciprocal space T-matrix results on the onsite potential of the defects; VII. Joint density of states (jDOS); VIII. Zoom-ins around the defects from Fig. 4 of the main text; IX. Details of the computation of the intervalley scattering map shown in Fig. 4c of the main text; X. Wavelength of the intervalley and intravalley patterns observed in Fig. 4 of the main text; XI. Boltzmann transport theory

**Author contributions**

F.J. discovered the native defects. F.J. carried out the STM measurements. Z.G., E. A. Q.-L., and F.J. fabricated the graphene on hBN heterostructures. C.B., S.P., and V.K performed the T-matrix calculations. F.J. carried out the tight-binding calculations. T.T. and K.W. synthesized the hBN crystals. F.J. and C.B. analyzed the data. A.F. carried out the theoretical electronic transport analysis. J.V.J. supervised the STM measurements and the sample fabrication. F.J., A.F, and J.V.J. wrote the manuscript, with input from all co-authors.

**Acknowledgments**

We thank David Goldhaber-Gordon and François Ducastelle for useful discussions.

**Funding Sources**

J.V.J. acknowledges support from the National Science Foundation under award DMR-1753367 and the Army Research Office under contract W911NF-17-1-0473. K.W. and T.T. acknowledge support from the Elemental Strategy Initiative conducted by the MEXT, Japan, Grant Number JPMXP0112101001 and JSPS KAKENHI Grant Number JP20H00354.

# Supplementary Material for : *Direct visualization of native defects in graphite and their effect on the electronic properties of Bernal-stacked bilayer graphene.*

Frédéric Joucken, Cristina Bena, Zhehao Ge, Eberth Quezada-Lopez, Sarah Pinon,
Vardan Kaladzhyan, Takashi Taniguchi, Kenji Watanabe, Aires Ferreira, Jairo Velasco Jr.

## CONTENTS

## I. ATOMIC-SCALE IMAGES OF DEFECTS IN EXFOLIATED GRAPHITE AND IN GRAPHENE DEVICES

We display in Fig. S1 representative STM images of defects encountered in exfoliated graphite and in graphene devices. The tip condition is not always optimum when imaging these defects because this is done after acquiring large-scale images (several hundreds of nm wide) at low tip-sample bias (a few mV), as explained in the main text. This is because large-scale low bias imaging allows localization of these defects thanks to the long range intravalley scattering pattern centered around them. Finding such defects randomly is very unlikely given their low concentration. Consequently, the image quality is not always optimal. However, we have encountered several defects in a bilayer graphene device which we identify as nitrogen dopants in the graphitic configuration [1], one of them is highlighted by a red frame in Fig. S1.

The identification of these defects as nitrogen dopants is based on their topographic signature [1] as well as on the comparison between the experimental tunneling spectrum acquired atop the defect and the computed LDOS (obtained with pybinding [3]) for the dopant, represented by an onsite energy of $-10$ eV, as shown in Fig. S2. Such onsite energy is consistent with a nitrogen dopant in monolayer graphene [4] (see also our recent work where we present gate-dependent spectroscopic data obtained on these native dopants [2]). This justifies our choice of this onsite energy for the T-matrix based calculations presented in the main text. For the tight-binding calculation, we used a circular bilayer graphene area with a radius of 100 nm with the dopant at the center. The dopant is modeled by changing the onsite energy of the sublattice site where it is lying (the neighbors are unchanged). The computation used the kernel polynomial method [5], as implemented in pybinding. A broadening width of 15 meV was used.

Finally, based on their topographic signature, we tentatively identify some defects encountered in exfoliated graphite as nitrogen dopants (indicated in Fig. S1). Further inquiry is however necessary for a definite identification of the native defects in these graphite samples.

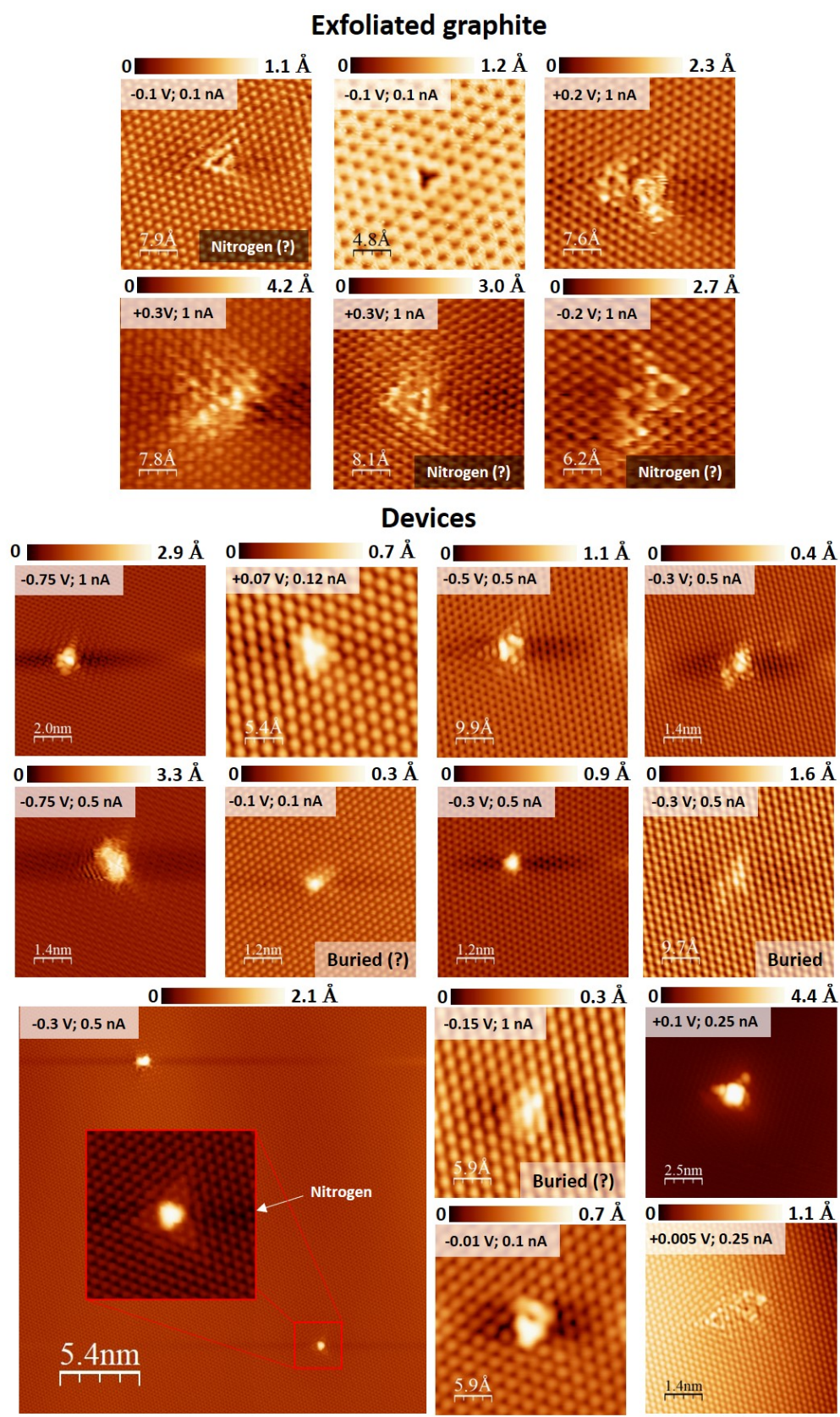

FIGURE S1 – Representative defects encountered on freshly cleaved graphite and on the graphene/hBN devices. Most defects are hardly identifiable because of bad tip states. However, we have encountered several times in the bilayer graphene devices a defect which we could identify as a nitrogen dopant (highlighted in red, see also Fig. S2 for STS data); see also our recent work where we present gate-dependent spectroscopic data obtained on these native dopants [2]. We tentatively identify some defects encountered in graphite as nitrogen (indicated by "Nitrogen (?)"). Some defects are clearly buried, whereas it is less clear in other cases (indicated by a question mark).

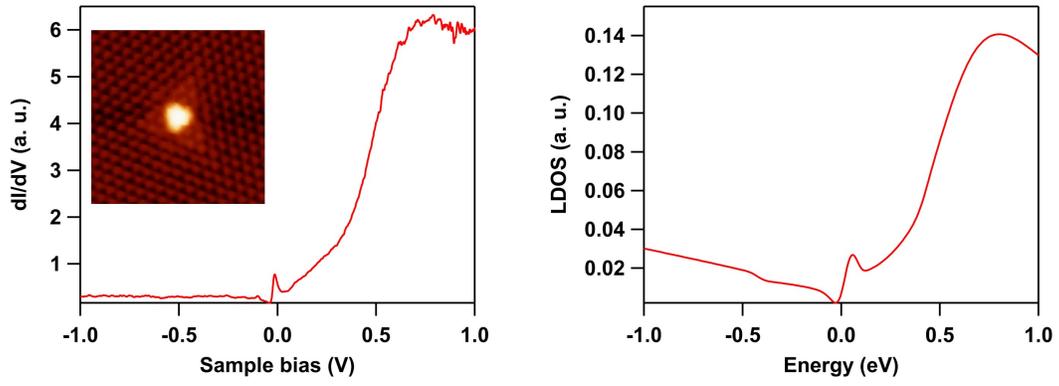

FIGURE S2 – (a) Experimental dI/dV spectrum obtained atop a nitrogen dopant shown in the inset ($3 \times 3$ nm$^2$, $-0.3$ V ; 0.5 nA). (b) Simulated LDOS for a dopant with an onsite potential of $-10$ eV.

## II. CONCENTRATION OF DEFECTS

We have measured the concentration of defects both on exfoliated graphite (exfoliated in situ) and on graphene devices. The graphene devices include two bilayer graphene samples and two thicker samples (5-6 and 10-11 Layers). On graphite, we have scanned a total area of 11,215,500 nm$^2$ and have found $\sim 75$ defects on the top layer in this area. This amounts to a concentration of $6.6 \times 10^8$ cm$^{-2}$. In graphene devices, we have scanned a total area of 1,922,400 nm$^2$ and have observed $\sim 32$ defects located in the top layer in this area. This amounts to a concentration of $2.2 \times 10^9$ cm$^{-2}$, significantly higher than in the graphite samples. This indicates that our fabrication process, which is commonly used by other researchers that study graphene in UHV, likely creates defects, in addition to the native defects already present in the parent crystal.

To evaluate the uncertainty on our estimation of the defects concentration, we computed the probability of observing $n$ defects, with the assumption that our measured concentration is the real concentration, and further assuming a random and uniform defect concentration. According to standard combinatorics and statistics, the probability $p(n)$ of observing n defects amongst $N$ atomic sites is given by $p = \binom{N}{n} c^n (1-c)^{N-n}$, where $N$ is the total number of atomic sites scanned and $c$ is the defect concentration. The function $p(n)$ is plotted in Fig. S3 for $N = 4.28 \times 10^8$ and $c = 1.71 \times 10^{-7}$, corresponding to the number of atomic sites we have scanned and the defects concentration [(# defects)/(# sites)] we have measured, respectively (this value of $N$ corresponds to the total scanned area of 11.2 $\mu m^2$). Considering a 95% interval and the width of the quasi-Gaussian distribution of $p(n)$, we estimate the uncertainty on the concentration to be about $\pm 2 \times 10^8$ cm$^{-2}$, as stated in the main text. The same procedure was applied to determine the uncertainty for the defect concentration in the graphene devices reported in the main text.

The probability of having zero defect as a function of the area presented in Fig. 1b of the main text is $(1-c)^N$, where $N$ corresponds to the number of atomic sites in the area and $c$ is the concentration we have measured ($c = 1.71 \times 10^{-7}$).

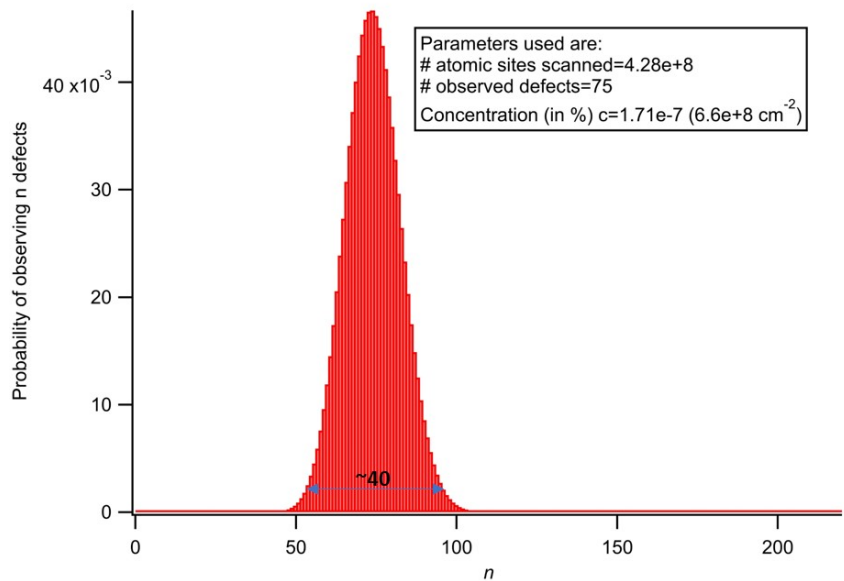

FIGURE S3 – To illustrate the error we make in our evaluation of the defect concentration, we plot here the probability distribution of observing $n$ defects, given the concentration of defects ($c = 1.71 \times 10^{-7}$) in the graphite top layer found in the total area that we have scanned (11.2 $\mu m^2$), assuming a random and homogenous distribution of the defects

## III. EXTRACTION OF THE GAP AND THE ELECTRONIC DOPING FOR THE BLG DEVICE FOR EACH GATE VOLTAGE

To extract the values of the electronic doping ($E_F - E_{CNP}$) and of the gap used in the T-Matrix simulation, we proceeded as follows. For the gap, we used a series of $dI/dV_S$ measurements at various gate voltages, presented in

Fig. S4a. The horizontal features in Fig. S4a correspond to the onset of phonon-assisted tunneling in the BLG/hBN sample, at $V_S \approx \pm 65$ meV [6, 7], which is gate independent. The phonon-assisted tunneling renders the visualization and determination of the gap problematic for energies above the inelastic tunneling threshold [8]. We thus determined the gap value (Fig. S4b) for a gate voltage where the gap can be measured precisely (15 V), and also determined the gate voltage where the gap closes (indicated by the red arrow, at $V_G \approx -52$ V), and extrapolated linearly between these two values. For the determination of the electronic doping, we determined the Fermi wavevector from the size of the experimental intravalley QPI pattern (Fig. 2 of the main text), similarly to what we recently reported [9]. The other tight-binding parameters used were $\gamma_0 = +3.3$ eV, $\gamma_1 = +0.42$ eV, and $\gamma_3 = -0.3$ eV [9, 10].

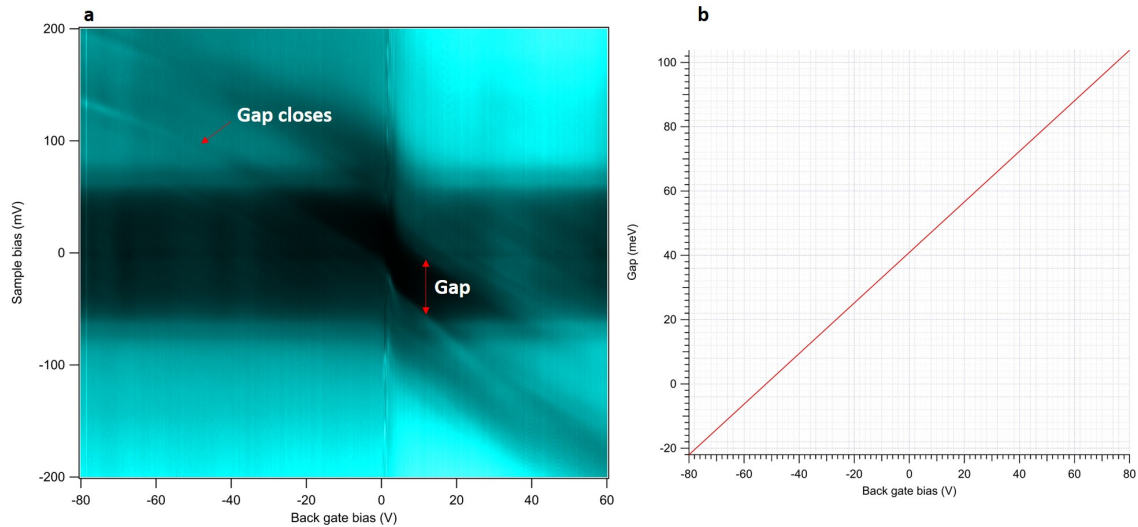

FIGURE S4 – Determination of the gap for the BLG/hBN device. (a) Plot of $dI/dV_S$ spectra as a function of gate voltage for the bilayer graphene sample used in Fig. 2, 3, and 5 of the main text. The gap is clearly visible for gate voltage around $V_G = 15$ V. The gap also clearly closes around $V_G = -52$ V. (b) Gap values linearly extrapolated from (a).

## IV. REAL SPACE IMAGING OF NATIVE DEFECTS IN BLG/hBN DEVICES

We present in this section several real-space intravalley scattering signatures induced by quasiparticles scattered off localized defects. Figures S5a and S5b show low-bias (5 mV) topographic STM images obtained at high tunneling current (20 nA), at opposite gate voltages ($V_G = -25$ V and $V_G = +25$ V, respectively). We found that the use of high current setpoints reveals the scattering patterns more clearly in topographic maps ; the same patterns are also visible in $dI/dV_S$ maps at lower current setpoints, as discussed in section V. Note that topographic STM images at low bias are essentially equivalent to spatial maps of LDOS($E_F$) [11–13]. Strikingly, a clear difference in contrast between the two gate polarities can be seen (Fig. S5a and S5b), with the positive gate (electron doping) inducing more pronounced intravalley scattering patterns, consistent with what is observed in Fourier space (see the discussion around Figs. 2 and 3 of the main text). We also find that the symmetry of the scattering pattern is triangular for two defects (top and bottom right of Fig.S5b), whereas it appears circularly symmetric for the other defects.

We have simulated the real-space signature for defects located at the four atomic sites of the Bernal-stacked bilayer graphene unit cell (inset of Fig. S5a). The defects are modeled by a variation of the onsite potential of $-10$ eV for a single atomic site (the choice of $-10$ eV is discussed in Section 1 and 6 of this document). The T-matrix results show that the triangular symmetry appears when the defect is in the bottom layer ($A_2$ or $B_2$), while features with a quasi-circular symmetry are observed for defects in the top layer ($A_1$ and $B_1$). Thus, these simple T-matrix calculations, which consider the modification of the onsite energy of a single atomic site, reproduce well the observed experimental patterns and can explain the two symmetries (circular and triangular) observed experimentally. This strongly indicates that the main factor determining the symmetry of the QPI patterns is the layer distribution of the defects. However, we found in experiments that some defects appearing in the top layer induce a triangular QPI pattern and some buried defects induce circular patterns. We attribute this to our simplified model that changes the onsite energy of a single atomic site, whereas the potential attributed to defects is likely more extended in real space [4].

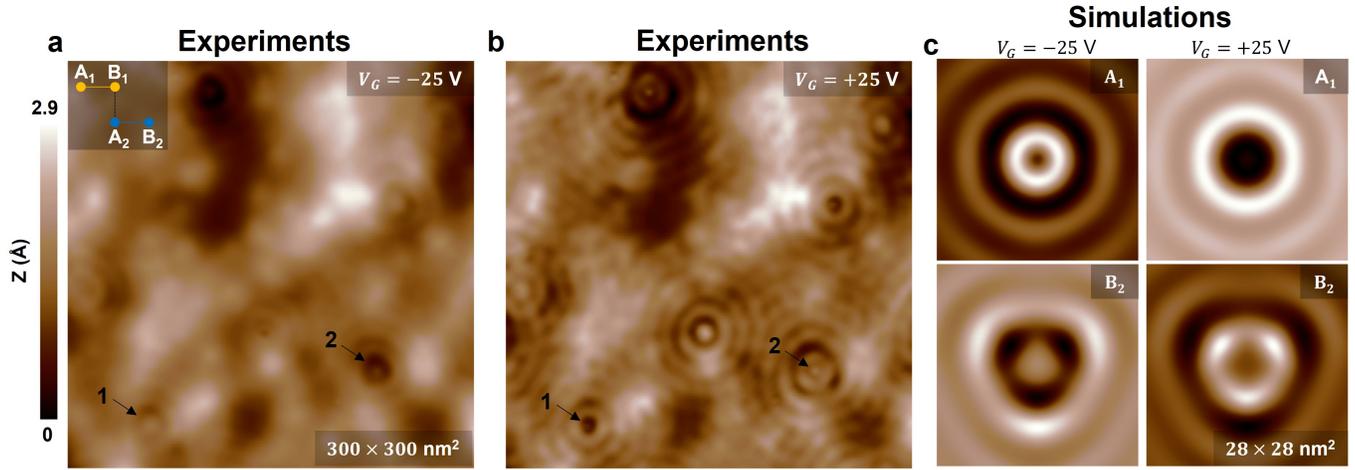

FIGURE S5 – QPI patterns induced by atomic-scale defects in Bernal-stacked bilayer graphene observed in real space. (a) Topographic low-bias (5 mV) STM image obtained at $V_G = -25$ V. (b) Same as in (a) at $V_G = +25$ V. Various QPI patterns evidently originating from localized sources can be observed. Some (as the one labelled "1") display quasi-circular symmetry whereas others (as the one labelled "2") display triangular symmetry. (c) T-matrix-based LDOS calculations for defects (onsite potential $V = -10$ eV) located in the top ($A_1$) or bottom ($B_2$) layer.

To support this hypothesis, we present real-space T-matrix simulation results for defects with more complex potentials in Fig. S6. Figure S6 displays real space simulation results for more complex defect types showing that top layer defects can also produce triangular QPI patterns. These features are obtained by summing up real space patterns obtained for a single dopant located on a particular sublattice site.

One can see that when the top layer dopant is on $B_1$, the overall pattern is triangular. That is not the case when the top layer dopant is on $A_1$. This is because the amplitude of the QPI pattern is in general significantly greater when the dopant is located on $A_1$, compared to $B_1$. This in turn can be linked to the much greater local density of states around the $A_1$ sublattice, compared to the $B_1$ sublattice, at low energy [2, 14, 15].

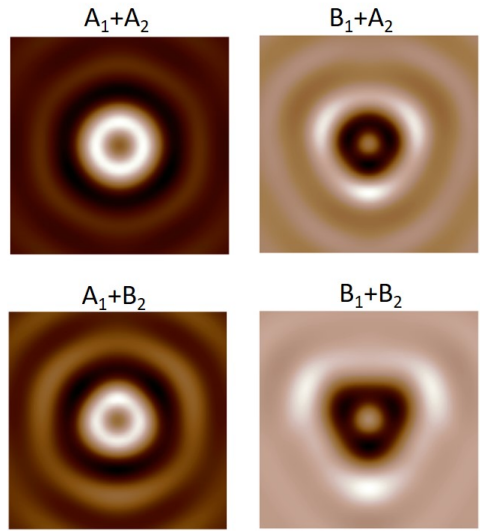

FIGURE S6 – Real space simulation results for more complex defect types showing that top layer defects can also produce triangular QPI patterns. The simulations are simply additions of the real space patterns obtained for a single dopant located on a particular sublattice site. One can see that when the top layer dopant is on $B_1$, the overall pattern is clearly triangular, which is less the case when the top layer dopant is located on $A_1$. This is due to the fact that the amplitude of the QPI pattern is in general significantly greater when the dopant is located on $A1$, compared to $B_1$. This in turn can be linked to the much greater local density of states around the $A_1$ sublattice, compared to the $B_1$ sublattice, at low energy [2, 14, 15].

## V. STM TOPOGRAPHIC MAP AND $dI/dV_S$ MAPS AT LOWER CURRENT

We show in Fig. S7 an STM topographic image (Fig. S7a) and a $dI/dV_S$ map (Fig. S7b) acquired at 1 nA and 0.25 nA (respectively) and displaying similar features as in Fig. S5, for which a current of 20 nA was used. This shows that working at high current (20 nA) is not required for observing these features. We noticed however that for certain tips, working at high current facilitates the observation of these features.

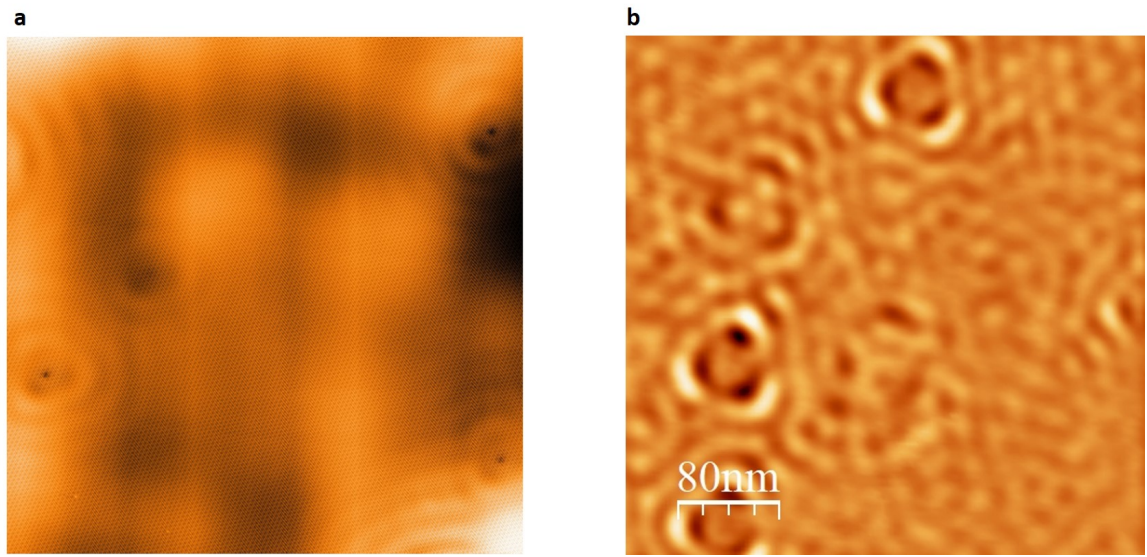

FIGURE S7 – (a) $300 \times 300$ nm$^2$ STM topographic map obtained at 1 nA, $V_S = +4$mV, $V_G = -10$ V. (b) $400 \times 400$ nm$^2$ $dI/dV_S$ map obtained at 0.25 nA, , $V_S = +5$ mV, 3 mV excitation, $V_G = -10$V. Similar features as those seen in Fig. S5 can be observed.

We further show in Fig. S8 that scanning at high current (20 nA) and low voltage (5 mV) does not create defects. Figure S8 shows two images obtained at $I = 20$ nA and $V_S = +5$ mV. The right image was obtained after having scanned the same area 4 times with the same parameters (only the backgate voltage was changed). One can see that the same defects can be found on both images.

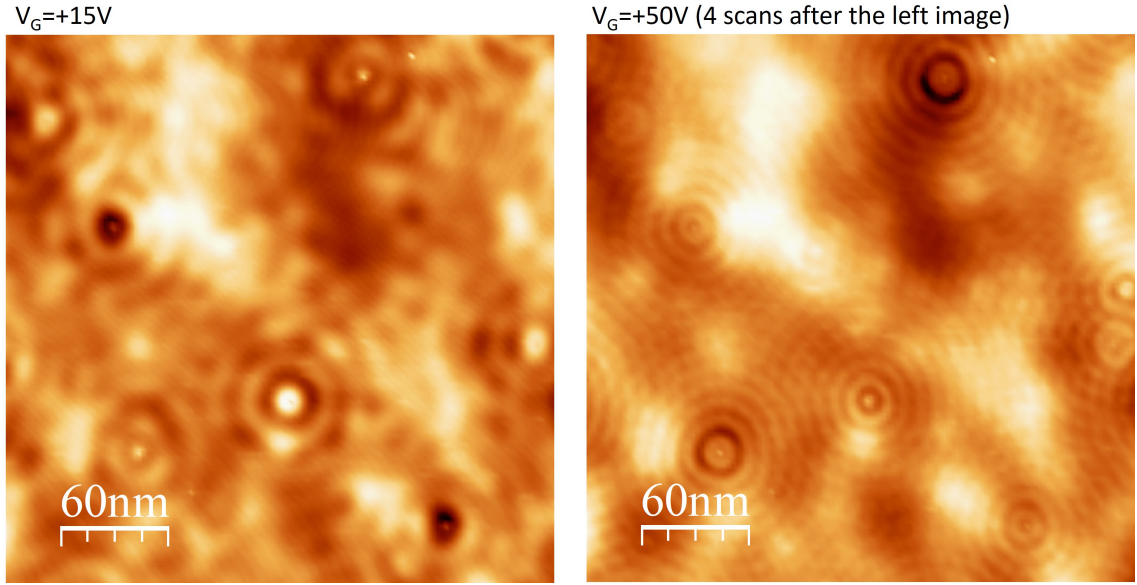

FIGURE S8 – Two images obtained at $I = 20$ nA and $V_S = +5$ mV. The right image was obtained after having scanned the same area 4 times with the same parameters (only the backgate voltage was changed). One can see that the same defects can be found on both images, demonstrating that the aggressive tunneling conditions do not create defects.

## VI. DEPENDENCE OF THE REAL-SPACE AND RECIPROCAL SPACE T-MATRIX RESULTS ON THE ONSITE POTENTIAL OF THE DEFECTS

We show in Fig. S9a the dependence of the real-space signature of the intravalley scattering on the onsite potential in T-matrix calculations. A significant increase in amplitude is seen as the onsite potential is increased (see z-scales). In Fig. S9b, we show the dependence of the intravalley pattern on the sublattice position. The amplitude of the scattering is clearly enhanced when the dopants is placed on the non-dimer sites ($A_1$ or $B_2$). This can be linked to the higher local density of states on these sites [2, 14, 15].

In Fig. S10, we present T-matrix results illustrating the dependence of the reciprocal scattering on the onsite potential. In Fig. S10a, we show T-matrix k-space QPI signature at two different energies (indicated) for two opposite signs of the onsite potential associated to the dopant ($\pm 10$ eV), averaged over the four sublattice sites. One can see that little difference is visible between the two cases. In (b) we show the T-matrix k-space QPI signature at two different energies (indicated) for three different onsite potential amplitude (indicated), averaged over the four sublattice sites. $V = -10^5$ eV corresponds to the vacancy case. The overall shape of the scattering patterns is not strongly influenced by the onsite potential energy. However, the amplitude in the FFT increases significantly with the onsite potential energy, reflecting the higher scattering rate for higher potential values. This also agrees with the real space results shown in Fig. S9.

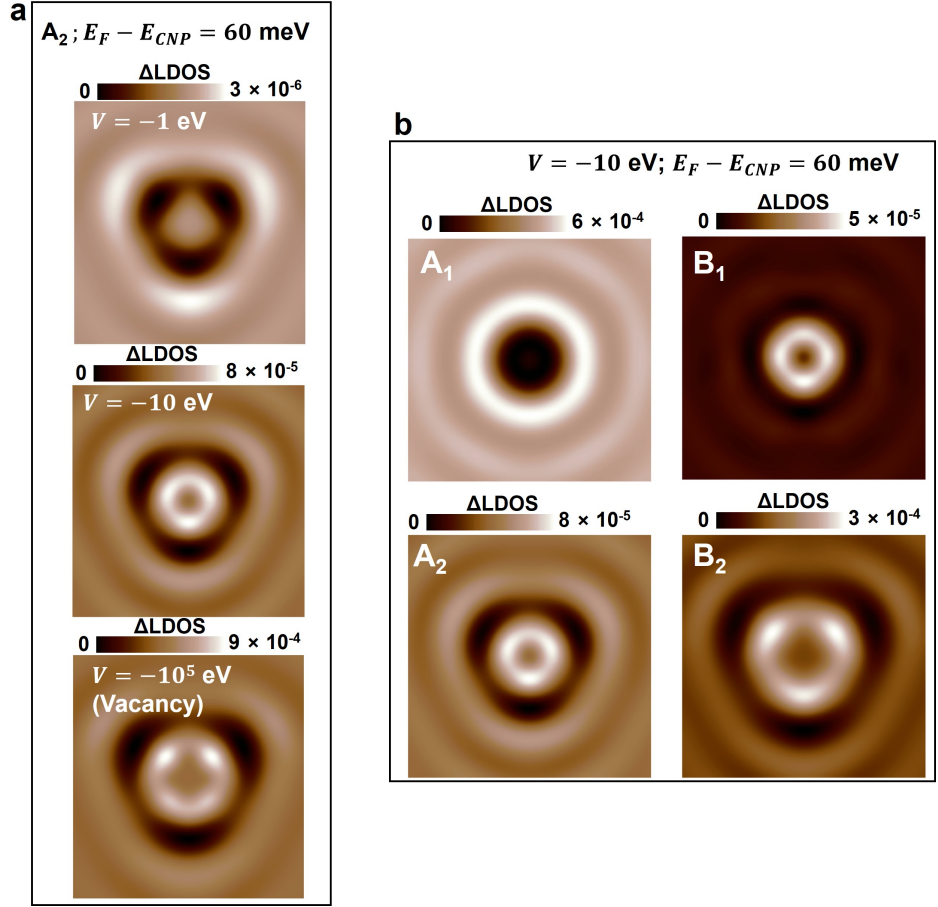

FIGURE S9 – (a) T-matrix based LDOS calculations for three values of the onsite potential ($V = -10^5$ eV corresponds to a vacancy). That shows A significant increase of the LDOS perturbation can be noticed as the onsite potential value is increased. In panel (b), we show the influence of the position of the dopant on the LDOS modulation. The LDOS variation is significantly greater when the dopant lies on the dimer sites ($A_1$ and $B_2$).

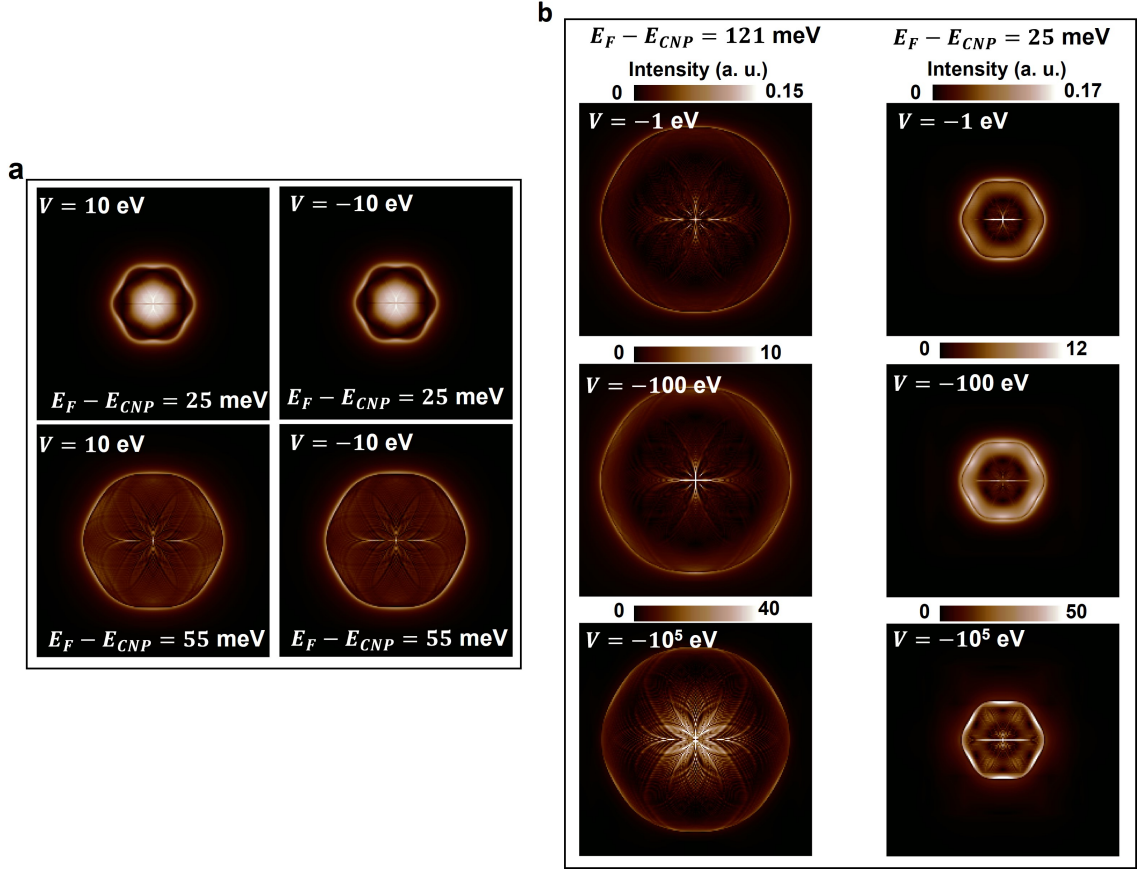

FIGURE S10 – In panel (a), we show T-matrix k-space QPI signature at two different energies (indicated) for two opposite signs of the onsite potential associated to the dopant ($\pm 10$ eV), averaged over the four sublattice sites. One can see that little difference is visible between the two cases. In (b) we show the T-matrix k-space QPI signature at two different energies (indicated) for three different onsite potential amplitude (indicated), averaged over the four sublattice sites. $V = -10^5$ eV corresponds to the vacancy case. The overall shape of the scattering patterns is not strongly influenced by the onsite potential energy. However, the amplitude in the FFT increases significantly with the onsite potential energy, reflecting the higher scattering rate for higher potential values. This also agrees with the real space results shown in the previous figure.

## VII. JOINT DENSITY OF STATES (JDOS)

In Fig. S11 we show two computed jDOS distributions at the energies corresponding to $V_G = +20$ V and $V_G = +60$ V (Fig. 2c and e and Fig. 3a and c of the main text; see our previous work for details [9]). One can see that the enhanced intensity for small wavevectors (small momentum distribution; see discussion around Figs. 2 and 3 in the main text) at low gate voltage is not reproduced in the jDOS. This shows that jDOS arguments cannot explain this feature.

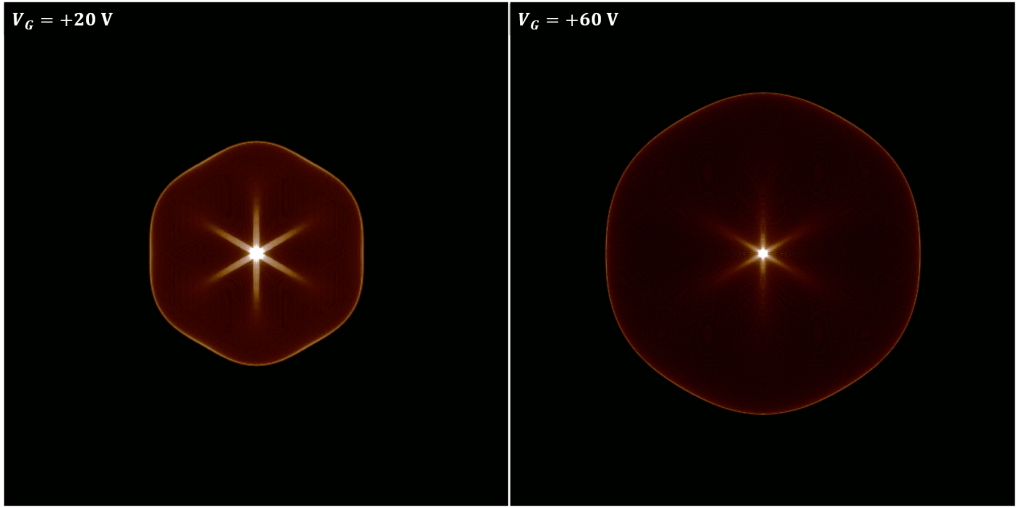

FIGURE S11 – jDOS computed for the experimental conditions at $V_G = +20$ V and $V_G = +60$ V (gap of 52 and 88 meV, and at $E_F - E_{CNP} = 52$ and 127 meV, respectively). One can see that jDOS considerations only do not reproduce the filled pattern seen in the experiments and in the T-matrix calculation.

## VIII. ZOOM-INS AROUND THE DEFECTS FROM FIG. 4 OF THE MAIN TEXT

Fig. S12 displays zoom-ins around the four atomic defects in the image of Fig. 4a of the main text. Defects number 1 and 2 are buried, whereas defects 3 and 4 are in the top layer.

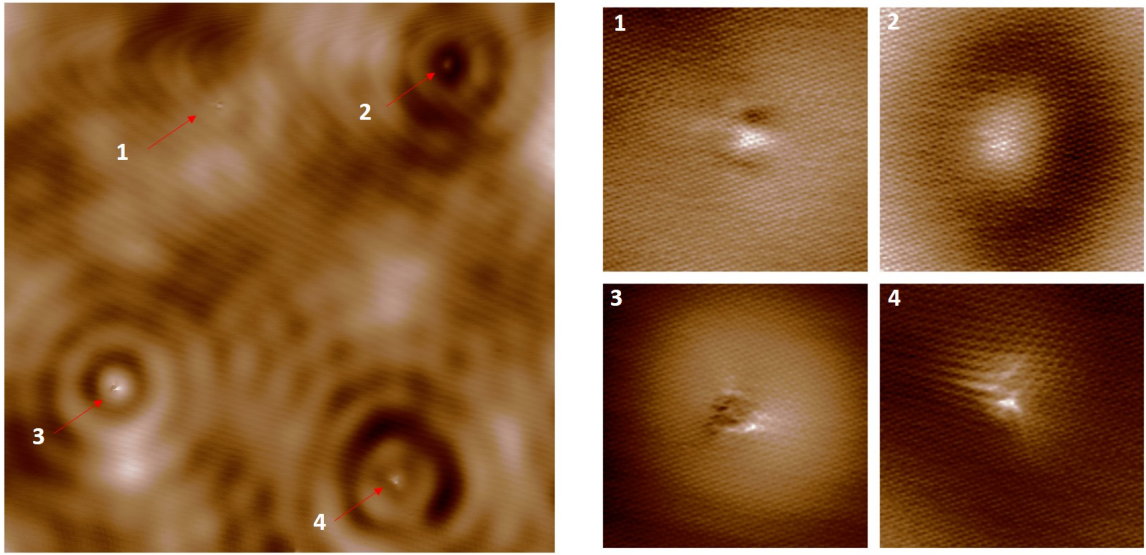

FIGURE S12 – Zoom-ins of the image in Fig. 4a of the main text to show the defects that are buried or on the top surface. 1 & 2 are buried; 3 & 4 are on the surface.

## IX. DETAILS OF THE COMPUTATION OF THE INTERVALLEY SCATTERING MAP SHOWN IN FIG. 4C OF THE MAIN TEXT

The details of the procedure followed to produce the intervalley scattering map presented in Fig. 4c of the main text is illustrated in Fig. S13. We start with the raw large scale image shown in Fig. S13a. It has $2048^2$ pixels for a size of $160^2$ nm$^2$. We then choose a pixel size for the zoom-ins [$32^2$ or $64^2$ pixels (corresponding to $2.5^2$ or $5^2$ nm$^2$) work well in this case], as well as a step size separating each zoom-in. Note that this step size can be smaller than the zoom-in size. One $64 \times 64$ pixels zoom-in is shown in Fig. S13b. For each zoom-in, we compute its FFT. The FFT of the zoom-in of Fig. S13b is shown in Fig. S13c. The intervalley scattering intensity for the corresponding spatial location of the zoom-in is then computed by summing the total intensity comprised within the red boxes in Fig. S13c, where the intervalley scattering peaks are expected (some intensity can be seen within the red boxes in Fig. S13c) and normalizing this sum by the sum of the intensity of the lattice Bragg peaks (boxed in white in Fig. S13c). The result is reproduced in Fig. S13d (same as Fig. 4c in the main text). A step size of 8 pixels (corresponding to 0.625 nm was used). Note that the normalization by the Bragg peak intensity is not strictly necessary but can be convenient for comparing absolute values of scattering intensity because ($i$) the FFT intensity depends on the zoom-in size and ($ii$) it can also correct for some tip change that can change the intensities of the peaks in the FFT.

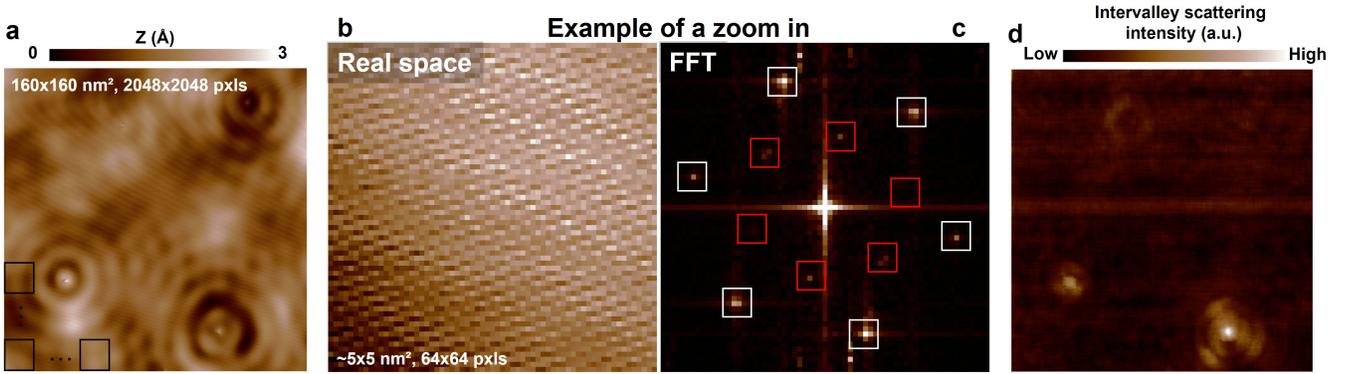

FIGURE S13 – Details of the computation of the intervalley scattering map shown in Fig. 4c of the main text. We start with the raw large scale image shown in panel a. It has $2048^2$ pixels for a size of $160^2$ nm$^2$. We then choose a pixel size for the zoom-ins ($32^2$ or $64^2$ pixels ($2.5^2$ or $5^2$ nm$^2$) work well in this case), as well as a step size separating each zoom-in. Note that this step size can be smaller than the zoom-in size. One $64 \times 64$ pixels zoom-in is shown in panel b. For each zoom-in, we compute its FFT. The FFT of the zoom-in of panel b is shown in panel c. The intervalley scattering intensity for the corresponding location of the zoom-in is then computed by summing the total intensity comprised within the red boxes in panel c, where the intervalley scattering peaks are expected (some intensity can be seen within the red boxes in panel c) and normalizing this sum by the sum of the intensity of the lattice Bragg peaks (boxed in white in panel c). The result is reproduced in panel d (same as Fig. 4c in the main text). A step size of 8 pixels (corresponding to 0.625 nm was used). Note that the normalization by the Bragg peak intensity is not strictly necessary but can be convenient for comparing absolute values of scattering intensity because (i) the FFT intensity depends on the zoom-in size and (ii) it can also correct for some tip change that can change the intensities of the peaks in the FFT.

## X. WAVELENGTH OF THE INTERVALLEY AND INTRAVALLEY PATTERNS OBSERVED IN FIG. 4 OF THE MAIN TEXT

In Fig. S14, we illustrate in 1D the reason why the wavelengths observed for the intervalley and the intravalley scattering differ by a factor of 2 (see the discussion around Fig. 4 of the main text). The left column of Fig. S14 illustrates the intravalley scattering case, where two wavevectors are opposite. The right column illustrates the intervalley scattering case, where the two wavevectors have the same orientation, but their absolute difference (0.2 m$^{-1}$) is the same as for the intravalley case. One can see that the resulting beating for the intervalley case has a wavelength twice shorter than for the intravalley case.

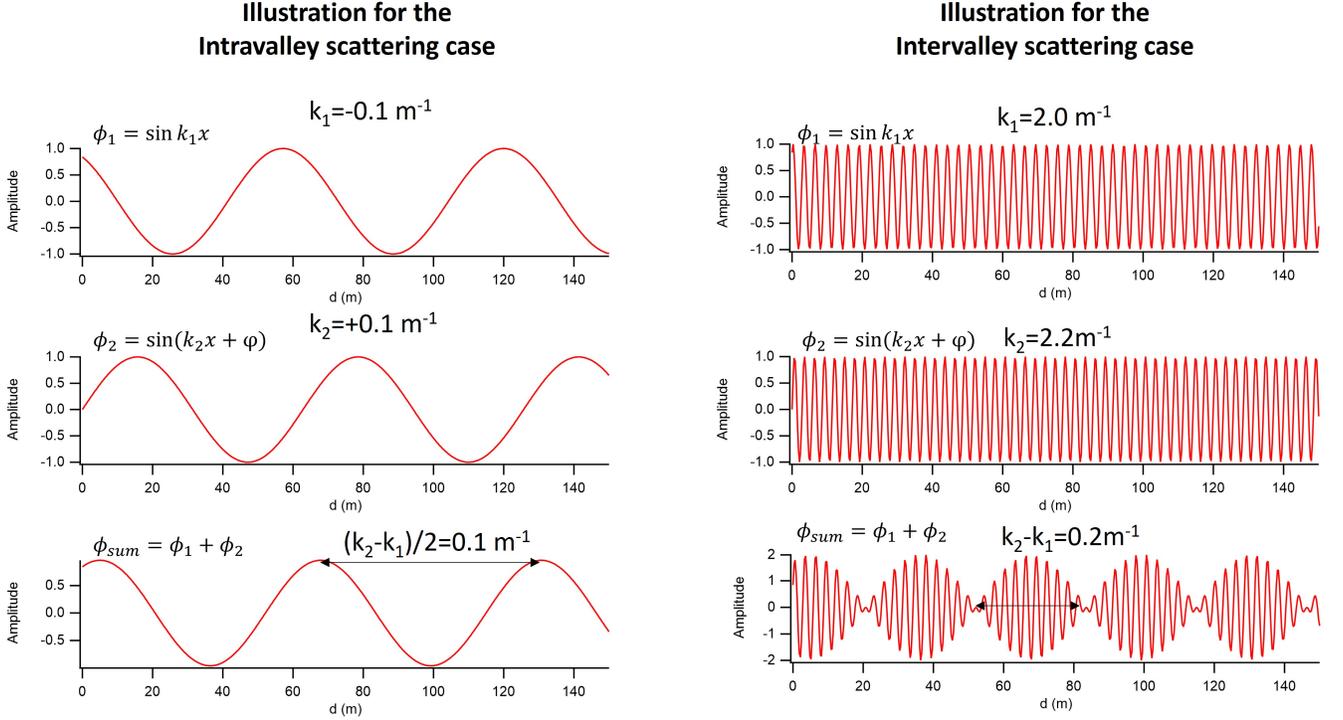

FIGURE S14 – Illustrating the reason why the wavelengths observed for the intervalley and the intravalley scattering differ by a factor of 2 (see the discussion around Fig. 4 of the main text). The left column illustrates the intravalley scattering case, where two wavevectors are opposite. The right column illustrates the intervalley scattering case, where the two wavevectors have the same orientation, but their absolute difference (0.2 m$^{-1}$) is the same as for the intravalley case. One can see that the resulting beating for the intervalley case has a wavelength twice shorter than for the intravalley case, as observed in Fig. 4 of the main text.

## XI. BOLTZMANN TRANSPORT THEORY

### A. Electronic structure

The $\pi$-electrons of Bernal-stacked bilayer graphene are governed by a Hamiltonian of four bands, but the low-energy physics ($|E| \ll |t_\perp|$, with $t_\perp$ the interlayer hopping between dimer carbon atoms $A_2$-$B_1$; see Fig. 3a in main text) can be approximated by an effective model of only two bands, where the dimer atoms linked by $t_\perp$ are projected out since they describe higher energy bands [16–19].

The low-energy Hamiltonian defined on the space of $A_1$-$B_2$ sites reads as

$$H_\tau = \frac{1}{t_\perp} \begin{bmatrix} 0 & \Pi_\tau^2 \\ \Pi_\tau^{\dagger 2} & 0 \end{bmatrix}, \qquad \Pi_\tau \equiv v(\tau p_x + i p_y) \quad (1)$$

where $\tau = \pm 1$ denotes the valley index and $v$ is the bare Dirac fermion velocity. Here, $t_\perp = 0.42$ eV and $v = 1.07 \times 10^6$ m/s [9, 10]. The dispersion relation obtained from Eq. (1) is easily computed to be

$$\varepsilon_{\lambda \tau \mathbf{p}} = \lambda \frac{v^2 p^2}{t_\perp}, \quad (2)$$

with $\lambda = \pm 1$ for electron (hole) states. As customary, the energies are measured with respect to the charge neutrality point of a pristine, half-filled system. The Bloch eigenvectors are also easily obtained :

$$|\psi_{\mathbf{k}}^{\lambda \tau}(\mathbf{r})\rangle = \frac{1}{\sqrt{2}} \begin{pmatrix} 1 \\ \lambda e^{-2i\tau \theta_\mathbf{k}} \end{pmatrix} e^{i \mathbf{k} \cdot \mathbf{r}}, \quad (3)$$

with $\theta_\mathbf{k} = \sphericalangle (\mathbf{k}, \hat{e}_x)$ the wavevector angle. We note that the Berry phase around each valley is $2\pi\tau$ (as opposed to the familiar value of $\pi\tau$ in monolayer graphene) and thus backscattering is allowed even for a smooth disorder potential.

The velocity operator $\mathbf{v}_\tau = (i/\hbar)[H_\tau, \mathbf{r}]$ reads as

$$\mathbf{v}_\tau = \frac{2v}{t_\perp} \begin{bmatrix} 0 & \Pi_\tau(\hat{e}_x + i\hat{e}_y) \\ \Pi_\tau^\dagger(\hat{e}_x - i\hat{e}_y) & 0 \end{bmatrix}. \tag{4}$$

Evaluating the matrix elements w.r.t. eigenvectors (3), one easily finds

$$\mathbf{v}_{\mathbf{k}\tau}^\lambda = \langle u_\mathbf{k}^{\lambda\tau}|\mathbf{v}_\tau|u_\mathbf{k}^{\lambda\tau}\rangle = \lambda \frac{2kv^2}{t_\perp}(\tau \cos\theta_\mathbf{k}\hat{e}_x + \sin\theta_\mathbf{k}\hat{e}_y), \tag{5}$$

with $k \equiv |\mathbf{k}|$. To find out the scattering transition rates for the bilayer system, we will also need the Green's function evaluated at the impurity position ($\mathbf{r} \equiv 0$). The first step is to derive the resolvent operator $G_\tau(z) \equiv (z - H_\tau)^{-1}$ of the 2-band Hamiltonian. Its Fourier transform satisfies the following equation

$$[z\sigma_0 - \beta|\mathbf{k}|^2 n_\tau(\mathbf{k})\cdot\boldsymbol{\sigma}] \cdot \mathcal{G}_\tau(z,\mathbf{k}) = \sigma_0, \tag{6}$$

where $\beta \equiv \hbar^2 v^2/t_\perp$ and $n_\tau(\mathbf{k}) \equiv [\cos 2\theta_\mathbf{k}, -\tau\sin 2\theta_\mathbf{k}]^\mathrm{T}$. Note that we have used the vector of Pauli matrices $\boldsymbol{\sigma}$ (supplemented with the identity operator $\sigma_0$) as a basis of the Clifford algebra. Inverting Eq. 6 is straightforward and yields

$$G_\tau(z,\mathbf{k}) = \frac{1}{z^2 - \beta^2\mathbf{k}^4}\left[z\sigma_0 + \beta\mathbf{k}^2 n_\tau(\mathbf{k})\cdot\boldsymbol{\sigma}\right]. \tag{7}$$

Performing the analytic continuation $z = \varepsilon + i0^+$, with $\varepsilon$ the Fermi energy, the required propagator evaluated at $\mathbf{r} = 0$ is obtained as

$$g_0(\varepsilon) \equiv \int \frac{d^2\mathbf{k}}{(2\pi)^2} G_\tau(\varepsilon + i0^+, \mathbf{k}) = \int_0^\infty \frac{dk}{2\pi} \frac{\varepsilon k}{\varepsilon^2 - \beta^2 k^4 + i0^+}. \tag{8}$$

This integral can be performed with standard methods and yields

$$g_0(\varepsilon) = \frac{i}{8\beta}. \tag{9}$$

### B. Scattering rates : T-matrix approach

The scattering rates are calculated using the generalized Fermi golden rule

$$W_{\mathbf{k}_\tau, \mathbf{k}'_{\tau'}} = \frac{2\pi}{\hbar} n_\mathrm{imp} \mathcal{T}_{\mathbf{k}_\tau, \mathbf{k}'_{\tau'}} \delta(\varepsilon_{\tau\mathbf{k}} - \varepsilon_{\tau'\mathbf{k}'}), \tag{10}$$

where $n_\mathrm{imp}$ is the areal density of short-range scatterers of a given type and

$$\mathcal{T}_{\mathbf{k}_\tau, \mathbf{k}'_{\tau'}} = |\langle \mathbf{k}'_{\tau'}|t|\mathbf{k}_\tau\rangle|^2, \qquad t \equiv t(\varepsilon) = \frac{\mathcal{V}}{1 - \mathcal{V}g_0(\varepsilon)}, \tag{11}$$

with $\mathcal{V}$ the Fourier transform of the impurity potential matrix. For top/bottom impurities (i.e. located on sublattice $A_1/B_2$), one has

$$\mathcal{V}_{A/B} = A_\hexagon^\mathrm{s.l.} M_{A/B}, \qquad M_{A/B} = \begin{bmatrix} u_A & 0 & u_A & 0 \\ 0 & u_B & 0 & u_B \\ u_A & 0 & u_A & 0 \\ 0 & u_B & 0 & u_B \end{bmatrix}. \tag{12}$$

with $A_\hexagon^\mathrm{s.l.}$ the monolayer unit cell area and $u_{A(B)}$ the onsite energy induced by an impurity on sublattice $A_1(B_2)$ [18]. The basis ordering is $A_1 K, B_2 K, A_1 K', B_2 K'$. The $t$ matrices for top/bottom impurities read as

$$t_A = \frac{A_\hexagon^\mathrm{s.l.} u_A}{1 - 2g_0 A_\hexagon^\mathrm{s.l.} u_A}\begin{bmatrix} 1 & 1 \\ 1 & 1 \end{bmatrix}, \qquad t_B = \frac{A_\hexagon^\mathrm{s.l.} u_B}{1 - 2g_0 A_\hexagon^\mathrm{s.l.} u_B}\begin{bmatrix} 1 & 1 \\ 1 & 1 \end{bmatrix}, \tag{13}$$

where only the relevant $2 \times 2$ sub-blocks are shown for simplicity. Thus,

$$\mathcal{T}^{A,B}_{\mathbf{k}_\tau,\mathbf{k}'_{\tau'}} = \frac{1}{4} \left| \frac{A^{\text{s.l.}}_\bigcirc u_{A,B}}{1 - 2g_0 A^{\text{s.l.}}_\bigcirc u_{A,B}} \right|^2. \tag{14}$$

Next, we evaluate the quasiparticle scattering rates. To ease the notation, we consider electron states ($\lambda = +1$) in the remainder of this note. We have

$$\Gamma^{A,B}_{\mathbf{k}\tau} = \sum_{\tau'} \int \frac{d^2\mathbf{k}'}{(2\pi)^2} W^{A,B}_{\mathbf{k}_\tau,\mathbf{k}'_{\tau'}} = \frac{2\pi}{\hbar} n^{A,B} \varrho_\varepsilon \int \frac{d\theta_{\mathbf{k}'}}{2\pi} \sum_{\tau'} \mathcal{T}^{A,B}_{\mathbf{k}_\tau,\mathbf{k}'_{\tau'}} \bigg|_{\varepsilon(k)=\varepsilon(k')} \tag{15}$$

where

$$\varrho_\varepsilon = \left( \frac{k}{2\pi\hbar|v_k|} \right)_{k=k(\varepsilon)} = \frac{1}{4\pi} \frac{t_\perp}{v^2 \hbar^2} \equiv \frac{1}{4\pi\beta} \tag{16}$$

is the density of states per valley/spin and $v_k = |\mathbf{v}_{\mathbf{k}\tau}|$ is the band velocity. Replacing Eq. (14) in Eq. (15), we obtain

$$\Gamma = \Gamma^A_{\mathbf{k}\tau} + \Gamma^B_{\mathbf{k}\tau} = \frac{1}{4\beta\hbar} n_{\text{dop}} \left| \frac{A^{\text{s.l.}}_\bigcirc u}{1 - 2g_0 A^{\text{s.l.}}_\bigcirc u} \right|^2, \tag{17}$$

where we assumed a uniform distribution of scatterers with $u_A = u_B = u$ and $n_{\text{dop}} = n^A + n^B$ is the total areal density of dopants.

## C. Longitudinal dc conductivity

The homogeneous Boltzmann transport equation (BTE) reads as

$$\frac{\partial f_{\mathbf{k}\tau}}{\partial t} + \dot{\mathbf{k}} \cdot \nabla_\mathbf{k} f_{\mathbf{k}\tau} = \frac{\partial f_{\mathbf{k}\tau}}{\partial t} \bigg|_{\text{collisions}}, \tag{18}$$

with $f_{\mathbf{k}\tau}$ the electron distribution function and $\dot{\mathbf{k}} = -e\mathbf{E}$ the force term in a dc electric field. To first order in $\mathbf{E}$, the steady-state BTE becomes

$$-e\mathbf{E} \cdot \mathbf{v}_{\mathbf{k}\tau} \left( \frac{\partial f^0_\mathbf{k}}{\partial \varepsilon} \right)_{\varepsilon=\varepsilon_{\tau\mathbf{k}}} = S[f_{\mathbf{k}\tau}], \tag{19}$$

where $f^0_\mathbf{k}$ is the equilibrium, Fermi-Dirac distribution function and

$$S[f_{\mathbf{k}\tau}] = \sum_{\tau'=\pm 1} \int \frac{d^2\mathbf{k}'}{4\pi^2} \left( f_{\mathbf{k}'\tau'} W_{\mathbf{k}'_{\tau'},\mathbf{k}_\tau} - f_{\mathbf{k}\tau} W_{\mathbf{k}_\tau,\mathbf{k}'_{\tau'}} \right) \tag{20}$$

is the scattering kernel.

The linearized BTE [Eq. (19)] for a system with isotropic Fermi surface has known exact solution

$$\delta f_{\mathbf{k}\tau} \equiv f_{\mathbf{k}\tau} - f^0_\mathbf{k} = (\mathbf{v}_{\mathbf{k}\tau} \cdot \mathbf{E}) \, e \, \tau_{\|\mathbf{k}\tau} \left( \frac{\partial f^0}{\partial \varepsilon} \right)_{\varepsilon=\varepsilon_{\tau\mathbf{k}}}, \tag{21}$$

with

$$\frac{1}{\tau_{\|\mathbf{k}\tau}} = \sum_{\tau'=\pm 1} \int \frac{d^2\mathbf{k}'}{4\pi^2} [1 - \cos(\theta_{\mathbf{k}'} - \theta_\mathbf{k})] W_{\mathbf{k}'_{\tau'},\mathbf{k}_\tau}. \tag{22}$$

Using the results derived earlier [Eqs. (10)-(14)], we obtain

$$\tau_\| \equiv \tau_{\|\mathbf{k}\tau} = \frac{4\beta\hbar}{n_{\text{dop}}} \left| \frac{1 - 2g_0 A_\bigcirc u}{A_\bigcirc u} \right|^2 = \frac{\hbar}{4 n_{\text{dop}} \beta} \left[ 1 + \left( \frac{8\beta}{A_\bigcirc u} \right)^2 \right], \tag{23}$$

with $A_{\hexagon} = 2A_{\hexagon}^{s.l.}$ the BLG unit cell area. Note that the mean free path is given by $l = v\tau_\|$.

Let us start by discussing the electronic transport properties in the zero-temperature limit. Without loss of generality, we set $\mathbf{E} = E_x \hat{x}$. The deviation of the distribution function at $T = 0$ acquires a particularly simple form

$$\delta f_{\mathbf{k}\tau} = -v_{k\tau,x} e \tau_\| \delta(\varepsilon_{\mathbf{k}\tau} - \varepsilon) E_x, \tag{24}$$

with $v_{\mathbf{k}\tau,x} = \frac{2\tau\hbar k_F v^2}{t_\perp} \cos\theta_{\mathbf{k}}$ [c.f. Eq. (5)].

The steady-state charge current density is given by

$$J_x = -e g_s \sum_{\tau=\pm 1} \int \frac{d^2\mathbf{k}}{(2\pi)^2} \, \delta f_{\mathbf{k}\tau} \, v_{\mathbf{k}\tau,x} \,, \tag{25}$$

with $g_s = 2$ the spin degeneracy factor. Plugging Eq. (24), we find

$$J_x = \frac{2e^2}{h} v_F k_F \tau_\| E_x \,, \qquad v_F \equiv \frac{2\hbar k_F v^2}{t_\perp},$$

The $T = 0$ dc-conductivity $\sigma_{xx} = J_x/E_x$ is now readily obtained

$$\sigma_{xx} = \frac{2e^2}{\pi} \frac{k_F^2 v^2}{t_\perp} \tau_\| = \frac{e^2}{h} \frac{k_F^2}{n_{\text{dop}}} \left(1 + \frac{64\beta^2}{A_{\hexagon}^2 u^2}\right). \tag{26}$$

The extrinsic electron mobility $\mu_e = \sigma_{xx}/(en_e)$ is found by expressing Eq. (26) in terms of the electronic density $n_e = \pi k_F^2$. For a typical impurity potential strength $u \approx \pm 10$ eV, one has

$$\mu_e \approx \frac{1}{\pi \hbar} \frac{e}{n_{\text{dop}}} \frac{64\beta^2}{A_{\hexagon}^2 u^2} \approx 6 \times 10^{15} \frac{1}{n_{\text{dop}}[\text{in cm}^2]} \, (\text{cm}^2/(\text{V.s})). \tag{27}$$

Next, we discuss the electron transport behavior at finite temperature. We neglect hydrodynamic effects [20], which is justified provided the chemical potential is not too low, i.e. $\mu \gtrsim k_B T$. We start by computing the thermal corrections to the impurity-limited conductivity. From Eqs. (21) and (25), we obtain

$$\sigma_{xx}(\mu, T) = \frac{e^2}{h} \left(1 + \frac{64\beta^2}{A_{\hexagon}^2 u^2}\right) \frac{\bar{\varepsilon}}{n_{\text{dop}}\beta} \,; \quad \bar{\varepsilon} \equiv \int_0^\infty d\varepsilon\, \varepsilon\, (-\partial_\varepsilon f) \,. \tag{28}$$

From this expression, we find $\bar{\varepsilon} \simeq \mu$ ($\mu \gg k_B T$) and thus, as expected, thermal fluctuations have little impact on the impurity contribution at finite carrier density [16–19]. We now discuss the role of electron-phonon collisions, which are expected to dominate the scattering rates at moderate-high temperatures. In Ref. [21], in-plane acoustic phonons were theoretically shown to provide the dominant contribution in strained bilayer samples. We note that the electron-phonon coupling in bilayer graphene has origin in scalar potential and synthetic gauge fields induced by strain deformation [21–24]. We confine our analysis to the latter contribution, which is known to dominate the acoustic-phonon resistivity in the low-temperature regime of interest to us [21, 24]. The corresponding conductivity is :

$$\sigma_{xx}^{\text{IP}} = \frac{e^2}{h} \frac{2\pi \rho v^2 k_B T}{8 k_F^2} \left\{ \sum_\nu \int_0^1 dx\, [D_B^\nu(2x)]^2 \frac{x^4}{\sqrt{1-x^2}} \frac{e^{xz_v}}{(e^{xz_v}-1)^2} \right\}^{-1}, \tag{29}$$

$$D_B^\nu(z) = \left[ 2g^2 y \left(1 - \frac{y^2}{2}\right)^2 \delta_{\nu L} + \frac{\hbar^2 v^2 \beta^2}{4a_0^2} \left(1 - \frac{y^2}{4}\right) \right]^{1/2} , \quad z_v \equiv \frac{2\hbar v_\nu k_F}{k_B T}, \tag{30}$$

where $g \approx 3$ eV is the effective deformation energy, $\rho \approx 7.6 \times 10^{-7}$ Kg/m$^2$ and $\beta = -\partial \log t / \partial \log a_0 \approx 3$, with $t$ the nearest-neighbor hopping and $a_0 \approx 0.14$ nm the carbon-carbon distance. For the phonon velocities, we take $v_L \approx 2.1 \times 10^4$ m/s, and $v_T \approx 1.4 \times 10^4$ m/s following the analysis of Ref. [21]. Equations (29)-(30) were used to calculate the temperature dependence of the phonon-limited conductivity in Fig. 5, main text.

---